\documentstyle[preprint,aps,prb,amssymb,amsfonts,verbatim,graphicx]{revtex}

\tightenlines

\begin{document}
\bibliographystyle{revtex}
\title{\bf A Pedestrian Introduction to Gamow Vectors}
\author{R.~de la Madrid}

\address{Institute for Scientific Interchange (ISI),
Villa Gualino, Viale Settimio Severo 65, I-10133, Torino, Italy \\
E-mail: \texttt{rafa@isiosf.isi.it}}
\author{M.~Gadella}
\address{Departamento de F\'\i sica Te\'orica, Facultad de Ciencias, 47011 
Valladolid, Spain}

\maketitle

\begin{abstract}
The Gamow vector description of resonances is compared with the 
$S$-matrix and the Green function descriptions using the example of
the square barrier potential. By imposing different
boundary conditions on the time independent Schr\"odinger equation,
we obtain either eigenvectors corresponding to real eigenvalues
and the physical spectrum or eigenvectors corresponding to complex
eigenvalues (Gamow vectors) and the resonance spectrum. We
show that the poles of the $S$ matrix are the same as the poles of
the Green function and are the complex eigenvalues of the
Schr\"odinger equation subject to a purely outgoing boundary
condition. The intrinsic time asymmetry of the
purely outgoing boundary condition is discussed. Finally, we show
that the probability of detecting the decay within a shell around
the origin of the decaying state follows an exponential law if the
Gamow vector (resonance) contribution to this probability is the
only contribution that is taken into account. 
\end{abstract}


\section{Introduction}
\setcounter{equation}{0}
\label{sec:introduction}
Most elementary particles are only quasistable states decaying 
through various interactions and thus have finite lifetimes of
various orders of magnitude.\cite{BARNETT} Several theoretical
descriptions have been proposed for quasistable states. Three of
the most widely used descriptions are the $S$ matrix, the
Gamow vector, and the Green function. We shall
use the example of nonrelativistic potential scattering,
in particular the square barrier potential, to demonstrate the
connection between these three descriptions.

Experimentally, resonances often appear as peaks in the cross
section that resemble the well-known Breit-Wigner distribution. The
Breit-Wigner distribution has two characteristic parameters: the energy
$E_R$ at which the peak reaches its maximum, and its width
$\Gamma_R$ at half-maximum. The inverse of 
$\Gamma_R$ is the lifetime of the decaying state.\cite{WW} The peak of
the cross section with Breit-Wigner shape is related to a first-order 
pole of the $S$ matrix in the energy representation $S(E)$ at the complex
number $z_R=E_R-i\Gamma_R /2$. The 
theoretical expression of the cross section in terms of $S(E)$
fits the shape of the experimental cross section in the neighborhood of 
$E_R$ (see, for example, Ref.~\onlinecite{BOHM}). This is why the
first-order pole of the $S$ matrix is often taken as the
theoretical definition of a resonance. The Green function description
treats a resonance as a pole of the Green function when
analytically continued to the whole complex plane. 
				
Although a resonance has a finite lifetime, it is otherwise assigned all the 
properties that are also attributed to stable particles, such as
angular momentum and charge. For
example, a radioactive nucleus has a finite lifetime, but otherwise
it possesses all the properties of stable nuclei; in fact, it is
included in the periodic table of the elements along with the
stable nuclei. Therefore, it seems natural to seek a theoretical
description that provides ``particle status'' to quasistable
states. The description of a resonance by Gamow vectors allows us
to interpret resonances as autonomous exponentially decaying
physical systems.

The energy eigenfunction with complex eigenvalue was
originally introduced by Gamow in his paper on $\alpha$ decay of
atomic nuclei,\cite{GAMOW} and used by a number of authors
(see, for example,
Refs.~\onlinecite{KPS,PEI,MONDRAGON,FERREIRA,BOLLINI,GASTON,BERGGREN}
and references therein). The real part of the complex eigenvalue is
associated with the energy of the resonance, and the imaginary
part is associated with the inverse of the lifetime. Gamow
eigenfunctions have an exponentially decaying time evolution, in
accordance with the exponential law observed in $\alpha$ decay of 
radioactive nuclei.\cite{ROLLEFSON,SUPON,DESMARAIS,RUDDICK}

Gamow's treatment was heuristic though, and could not be made 
mathematically rigorous within the Hilbert space theory, because
self-adjoint operators on a Hilbert space can only have real
eigenvalues. A rigorous mathematical treatment of Gamow vectors
needs an extension of Hilbert space to the Rigged
Hilbert Space (RHS).\cite{DIS,ANTOINE98,BG} The RHS was first
introduced in physics in order to justify Dirac's bra-ket
formalism.\cite{RHS1,RHS2,RHS3} In RHS language, Gamow vectors are
eigenkets of a (dual) extension of the self-adjoint Hamiltonian.
This extension can surely have complex eigenvalues. Using the RHS
formalism, one can prove that the time evolution of the Gamow
vectors is governed by a semigroup, expressing time asymmetry on 
the microscopic level. One
can also obtain an exact golden rule\cite{BOHM} for the
decay of the state described by the Gamow vectors. This golden rule
reduces to the Fermi-Dirac golden rule in the Born approximation. 

There are several pedagogical papers on Gamow
vectors\cite{BGM,HOLSTEIN95} and related topics such as $\alpha$
decay,\cite{HOLSTEIN95,FUDA} barrier 
penetration,\cite{HOLSTEIN83,MASSMANN} and exponential 
decay\cite{HOLSTEIN83,MASSMANN,LEVITAN,ONLEY} (this list is not
exhaustive). Some of these papers consider alternatives to the
Gamow, $S$ matrix, and Green function descriptions. Especially 
suggestive is the approach of Refs.~\onlinecite{HOLSTEIN95} and
\onlinecite{HOLSTEIN83}, where among other methods Feynman's path
integral is used. In this paper, we discuss at the graduate student
level some of the issues not covered by
Refs.~\onlinecite{BGM,HOLSTEIN95,FUDA,HOLSTEIN83,MASSMANN,LEVITAN,ONLEY}.

In Sec.~\ref{sec:time}, we calculate the Dirac kets for the square
barrier potential. These Dirac kets are monoenergetic
eigensolutions of the time independent Schr\"odinger equation.
They are not square normalizable, and therefore they cannot
represent a wave packet. Actually, they are members of a
continuous basis that expands the wave functions in the
Dirac basis vector expansion.

The well-known expression for the $S$ matrix in the energy
representation is provided in Sec.~\ref{sec:s-matrix}. The
resonances will be defined as the poles of the $S$ matrix. In
Sec.~\ref{sec:gamow}, we calculate the Gamow vectors as the
solutions of the time independent Schr\"odinger equation with
complex eigenvalues subject to purely outgoing boundary conditions.

The Green function and its poles are calculated in
Sec.~\ref{sec:green}. It will be apparent that the poles of the
Green function are the same as the poles of the $S$ matrix and are
the complex eigenvalues obtained from the purely outgoing
boundary condition. The residue of the Green function 
at the resonance energy is expressed in terms of the Gamow 
eigenfunction. 

In Sec.~\ref{sec:complex}, the basis vector expansion generated by
the Gamow vectors is provided. Sections~\ref{sec:outgoing}
and~\ref{Ovouiampi} discuss the time asymmetry built into the
purely outgoing boundary condition. Section~\ref{Compuprova}
treats the exponential decay law of the Gamow vectors.

In the paper, there will be some mathematical interludes enclosed
by brackets. These interludes are not essential to understand the
paper (although they are essential to prove other fundamental
results\cite{DIS,BG}) and may be skipped in the first reading. They
are recommended to readers who are
seriously interested in the mathematical framework that supports
the Gamow vector approach to resonance scattering.

\section{Resonances for a Square Barrier Potential}
\label{sec:resonances}

\subsection{Time Independent Schr\"odinger Equation}
\label{sec:time}

We consider a three-dimensional square barrier potential of
height $V_0>0$ and calculate the energy kets using the time
independent Schr\"odinger equation
\begin{equation}
	H|E\rangle =E|E\rangle \, . 
 \label{tiSe}
\end{equation}
Equation~(\ref{tiSe}) is an eigenvalue equation of the Hamiltonian
$H$. When the eigenvalues $E$ belong to the continuous spectrum of
$H$, then the solutions $|E\rangle$, also called {\it Dirac kets}, 
are given by vectors that lie outside the Hilbert space, that is,
they are not square integrable. Following von Neumann,
these kets are sometimes interpreted as states that are ``very
near a proper state.''\cite{VON} Following Dirac,\cite{DIRAC} we
interpret these kets as members of a continuous basis system
that spans the space of wave functions. Therefore, every wave function 
$\varphi$ can be expanded in terms of the Dirac kets 
$|E,l,m\rangle \equiv |E\rangle$ as
\begin{equation}
 \varphi =\sum_{l=0}^{\infty}\sum_{m=-l}^{l}\!\int_0^{\infty} \!
 dE \, |E,l,m \rangle \langle E,l,m|\varphi \rangle \, .
	\label{ve}
\end{equation}
[Although Dirac introduced Eq.~(\ref{ve}) on heuristic
grounds, its mathematical rigor was later established by the
Gelfand-Maurin theorem\cite{GELFAND} (also called the nuclear
spectral theorem) within the Rigged Hilbert Space (see also
Refs.~\onlinecite{DIS,BG,RHS1,RHS2,RHS3,FP}, and
\onlinecite{JPA}).]

To calculate the possible set of (real)
eigenvalues and their corresponding eigenvectors in our example, we
solve Eq.~(\ref{tiSe}) in the position representation, 
\begin{equation}
	\label{delta}
	\langle \vec{x}|H|E\rangle =
	\left(\frac{-\hbar^2}{2m} \nabla^2 +V(\vec{x})\right)
	\langle \vec{x}|E\rangle =E\langle \vec{x}|E\rangle \, ,
\end{equation}
where $\nabla^2$ is the three-dimensional Laplacian and
\begin{equation}
	V(\vec{x})=V(r)=\left\{ \begin{array}{ll}
 0 &0<r<a \\
 V_0 &a<r<b \\
 0  &b<r<\infty \, . 
 \end{array} 
 \right. 
	\label{potential}
\end{equation}
Because $V(\vec{x})$ is spherically symmetric, we can use
spherical coordinates $\vec{x}\equiv (r,\theta, \phi)$ to solve
Eq.~(\ref{delta}), which in
spherical coordinates reads 
\begin{equation}
	\langle r,\theta, \phi|H|E,l,m \rangle =
	\biggl(\frac{-\hbar^2}{2m}\frac{1}{r}\frac{\partial^2}{\partial
r^2}r
	+\frac{\hbar^2l(l+1)}{2mr^2}+V(r)\biggr)
 \langle r,\theta, \phi|E,l,m\rangle 
 = E\langle r,\theta, \phi|E,l,m\rangle\,.
	\label{sphSe}
\end{equation}
By separating the radial and angular dependences, 
\begin{equation}
	\langle r,\theta, \phi|E,l,m\rangle \equiv
	\langle r|E\rangle_l \, \langle \theta, \phi |l,m\rangle \equiv
	\frac{1}{r}\chi_l(r;E)Y_{l,m}(\theta, \phi),
\end{equation}
where $Y_{l,m}(\theta, \phi)$ are the spherical harmonics, we obtain for 
the radial part
\begin{equation}
\biggl(\frac{-\hbar^2}{2m}\frac{d^2}{dr^2}+\frac{\hbar^2l(l+1)}{2mr^2}
	+V(r) \biggr) \chi_l(r;E)=E\chi_l(r;E)\,.
	\label{baba}
\end{equation}
In this section, we shall restrict ourselves to the case of zero orbital 
angular momentum (the higher-order case is treated in the
appendix). We write 
$\chi_{l=0}(r;E)\equiv \chi (r;E)$ and obtain
\begin{equation}
	-\frac{\hbar^2}{2m} \frac{d^2}{dr^2}\chi(r;E)+V(r)\chi(r;E)=
 E\chi (r;E)\,.
	\label{rSe0}
\end{equation}
The solutions of Eq.~(\ref{rSe0}) for each value of $E$ are the kets
(more precisely, the radial part of the energy kets in the position
representation 
$\langle \vec{x}|E\rangle =\langle r,\theta , \phi|E,0,0 \rangle 
=\frac{\chi (r;E)}{r} Y_{0,0}(\theta, \phi)\,$). In our case,
\begin{equation}
 \chi (r;E)=\left\{ \begin{array}{ll}
 \alpha_1e^{ikr}+\beta_1e^{-ikr} &0<r<a \\
 \alpha_2e^{iQr}+\beta_2e^{-iQr} &a<r<b \\
 {\cal F}_1e^{ikr}+{\cal F}_2e^{-ikr} 
 &b<r<\infty \, ,
 \end{array} 
\right.
\label{rs}
\end{equation}
where 
\begin{equation}
	k=\sqrt{\frac{2m}{\hbar^2}E} 
 \label{momentum}
\end{equation}
is the wave number of the particle, and
\begin{equation}
	Q=\sqrt{k^2-\frac{2m}{\hbar^2}V_0}=\sqrt{\frac{2m}{\hbar^2}(E-V_0)}
 \, .
 \label{Qmomentum}
\end{equation}
The coefficients $\alpha$, $\beta$, and ${\cal F}$ are
functions of
$k$ (and therefore of the energy $E$). Their possible values are
restricted by the boundary conditions that we need to impose on
the solutions~(\ref{rs}) of the radial Schr\"odinger equation.
Some of these boundary conditions are due to the fact that at
least the second derivative of $\chi$ must be well
defined, which is why $\chi$ must be a continuous function of $r$
with a continuous derivative. Another boundary condition makes the
solution vanish at the origin (see Eq.~(\ref{jaja})). This
condition can be understood by viewing the analogous
one-dimensional problem of Eq.~(\ref{rSe0}). In the
one-dimensional case the potential is infinite for $r<0$. The
function ${\cal F}_2e^{-ikr}$ represents an incoming probability
wave of amplitude ${\cal F}_2$, and ${\cal F}_1e^{ikr}$ 
represents an outgoing wave of amplitude ${\cal F}_1$. Because
there cannot be any transmission into the region $r<0$,
$\chi$ must vanish at the origin.\cite{COHEN} Mathematically,
this condition is related to the self-adjointness of the
Hamiltonian.\cite{DUNFORD,DIS} Finally, the eigenfunction
$\chi(r;E)$ is assumed to be bounded (see Eq.~(\ref{bc})). We
impose this boundedness condition because we want the set of 
eigenvalues to coincide with the Hilbert space (that is,
the physical) spectrum of the Hamiltonian. In our case, the
Hilbert space spectrum is $[0, \infty )$ (see
Refs.~\onlinecite{DUNFORD} and \onlinecite{DIS}). Then, the
boundary conditions that we impose upon Eq.~(\ref{rs}) read
\begin{mathletters}
 \label{boucodi}
\begin{eqnarray}
	\chi (0;E)&=&0 \label{jaja} \\
	\chi (a-0;E)&=&\chi (a+0;E) \label{Dirboc2} \\
	\chi '(a-0;E)&=&\chi '(a+0;E) \\
	\chi (b-0;E)&=&\chi (b+0;E) \\
	\chi '(b-0;E)&=&\chi '(b+0;E) \label{pretlst} \\
	|\chi (r;E)|&<& \infty \, . \label{bc} 
\end{eqnarray} 
\end{mathletters} 
If $E$ is a negative real number (or complex), then there are no 
coefficients of $\chi (r;E)$ in (\ref{rs}) for which the boundary
conditions (\ref{boucodi}) can be satisfied unless they are
trivially zero. To be more precise, if $E$ is a complex or a
negative real number, the corresponding eigensolution $\chi (r;E)$
of Eq.~(\ref{rSe0}) does not satisfy the boundary condition
(\ref{bc}), even though it satisfies the other boundary
conditions (\ref{jaja})--(\ref{pretlst}). If $E$ is a real number
in 
$[0,\infty )$, the corresponding eigenfunction satisfies all the
boundary conditions in~(\ref{boucodi}). Therefore, the boundedness
of the eigenkets is what forbids the negative (and the complex)
energies and singles out the physical (Hilbert space) spectrum.

If we use the notation $\alpha =2i\alpha_1$, we rewrite the boundary
conditions~(\ref{boucodi}) as the following
equations for the coefficients:
\begin{mathletters}
\begin{eqnarray}
	 \alpha_2e^{iQa}+\beta_2e^{-iQa}&=&\alpha \sin (ka) \\
	iQ(\alpha_2e^{iQa}-\beta_2e^{-iQa})&=&\alpha k\cos (ka) \\
	{\cal F}_1e^{ikb}+{\cal F}_2e^{-ikb}&=&\alpha_2e^{iQb}
	+\beta_2e^{-iQb} \\
	ik({\cal F}_1e^{ikb}-{\cal F}_2e^{-ikb})&=&iQ(\alpha_2e^{iQb}-
	\beta_2e^{-iQb}) \, .
\end{eqnarray}
\end{mathletters} 
After straightforward but tedious calculations, we find that
\begin{mathletters}
 \label{coeff}
\begin{eqnarray}
	\alpha_2(k)&=&
	\frac{1}{2}e^{-iQa}\,\biggl [\sin (ka)+\frac{k}{iQ}\cos
(ka)\biggr ] \alpha (k)
	\label{coef1}\\
	\beta_2(k)&=&
	\frac{1}{2}e^{iQa}\,\biggl[\sin (ka)-\frac{k}{iQ}\cos (ka)\biggr]\,
 \alpha (k) \\
	{\cal F}_1(k)&=&\frac{e^{-ikb}}{4}
	\biggl[(1+\frac{Q}{k})e^{iQ(b-a)}
	(\sin (ka)+\frac{k}{iQ}\cos (ka)) \nonumber \\
&&{}+ (1-\frac{Q}{k})e^{-iQ(b-a)}
	(\sin (ka)-\frac{k}{iQ}\cos (ka))\biggr]\alpha(k) \label{F-0} \\
	{\cal F}_2(k)&=&\frac{e^{ikb}}{4}\biggl[(1-\frac{Q}{k})e^{iQ(b-a)}
	(\sin (ka)+\frac{k}{iQ}\cos (ka)) \nonumber \\
&&{}+(1+\frac{Q}{k})e^{-iQ(b-a)}
	(\sin (ka)-\frac{k}{iQ}\cos (ka))\biggr]\alpha(k) 
 \label{F+0} \, . 
\end{eqnarray}
\end{mathletters} 
Thus, for each $E$ in $[0,\infty)$, there exists a solution of the
eigenvalue equation in the position representation 
\begin{equation}
	\langle r,\theta, \phi |E\rangle =
 \frac{\chi (r;E)}{r}\, Y_{0,0}(\theta ,\phi) =
 \frac{\chi (r;E)}{r}\, \sqrt{ \frac{1}{4\pi } } \, , 
 \qquad 0\leq E<\infty
 \label{resul21}
\end{equation}
with $\chi (r;E)$ given by
\begin{equation}
 	r\langle r|E \rangle_{l=0}=
 \chi (r;E)=\left\{ \begin{array}{ll}
 \alpha (k) \sin (kr) &0<r<a \\
 \alpha_2(k)e^{iQr}+\beta_2(k)e^{-iQr} &a<r<b \\
 {\cal F}_1(k)e^{ikr}+{\cal F}_2(k)e^{-ikr} 
 &b<r<\infty \, .
 \end{array} 
 \right. 
	\label{sss0}
\end{equation}
The coefficients $\alpha_2(k)$, $\beta_2(k)$, and ${\cal F}_{1,2}(k)$
in Eq.~(\ref{sss0}) are given by Eq.~(\ref{coeff}) in terms of 
$\alpha (k)$. Usually, $\alpha (k)$
is chosen such that the kets in Eq.~(\ref{sss0}) are 
$\delta$-normalized, although in the following sections we shall
assume that
$\alpha (k)=1$. 

Equation (\ref{resul21}) means that for each energy eigenvalue $E$ in the 
spectrum $[0,\infty )$ of the Hamiltonian, there exists an 
eigensolution of the Hamiltonian satisfying the boundary conditions
(\ref{boucodi}). We can expand any wave function $\varphi$ 
in terms of these eigenfunctions (for $l=0$) as
\begin{equation}
	\varphi (r,\theta ,\phi )=\!\int_0^{\infty} \! dE \frac{\chi
(r;E)}{r} Y_{0,0}(\theta ,\phi)\,
 \varphi(E)\, ,
 \label{eigsosasdisr}
\end{equation}
or in bra-ket notation
\begin{equation}
	\langle r,\theta,\phi |\varphi \rangle =
\!\int_0^{\infty} \! dE\, \langle r,\theta,\phi |E\rangle 
 \langle E|\varphi \rangle \, .
	\label{vep0}
\end{equation}
We can interpret~(\ref{vep0}) by saying that any wave function 
$\varphi$ is a continuous linear superposition of the 
eigenfunctions $\langle r,\theta,\phi |E\rangle$ in
Eq.~(\ref{resul21}). Each eigenfunction is
weighted by $\langle E|\varphi \rangle$, which represents the wave
function in the energy representation. 

[Mathematically, the eigenkets $\langle r,\theta ,\phi |E\rangle$
allow us to go from the energy representation $\varphi (E)=\langle
E|\varphi \rangle$ to the position representation 
$\varphi (\vec{x})\equiv \langle r,\theta ,\phi |\varphi \rangle$
and vice versa, that is, they are continuous transition matrix
elements. Because the monoenergetic $\langle r,\theta,\phi
|E\rangle$ are not square integrable, they are not in the Hilbert
space. Thus Hilbert space methods are not sufficient to handle
them, and an extension of those methods is needed. As shown in
Refs.~\onlinecite{DIS,BG,FP}, and \onlinecite{JPA}, the extension
that seems to be the most convenient is the Rigged Hilbert Space.] 

\subsection{S-Matrix Approach}
\label{sec:s-matrix}
We consider now the scattering process of a particle beam off the
square barrier potential. The $S$ matrix for this process relates
the incoming (or prepared) and the outgoing (or detected) beams.
We write the $S$ matrix in the energy and angular momentum
representation,
\begin{equation}
 \langle {^-} E,l,m|E',l',m' {^+} \rangle =
 S_l(E)\, \delta (E-E')\, \delta_{l,l'}\, \delta_{m,m'} \, ,
\end{equation}
where $|E,l,m^{\pm} \rangle$ are the kets that solve the
Lippmann-Schwinger equation, and $\langle ^{\pm}E,l,m|$ are their corresponding
bras. In the position
representation, these Lippmann-Schwinger eigenfunctions are given
by (see Refs.~\onlinecite{NEWTON,TAYLOR,KUKULIN})
\begin{mathletters}
\begin{eqnarray}
 && \langle r|E^+\rangle =\frac{-1}{2i} \frac{\chi (r;E)}{ {\cal F}_2} \\
 && \langle r|E^-\rangle =\frac{1}{2i} \frac{\chi (r;E)}{ {\cal
F}_1} .
 \label{-LSchwingerke}
\end{eqnarray}
\end{mathletters}
The probability amplitude of detecting an out-state $\psi^-$ in 
an in-state $\varphi^+$ is (see Refs.~\onlinecite{BOHM} and
\onlinecite{DIS})
\begin{equation}
	(\psi^-,\varphi^+)=\sum_{l=0}^{\infty}
	\sum_{m=-l}^{l} \int_0^{\infty} ÆdE
 \langle \psi^-|E,l,m^-\rangle S_l(E)
 \langle^+ E,l,m |\varphi^+\rangle\, .
 \label{+-equat}
\end{equation}
In this section, we restrict ourselves to the case of zero angular
momentum, $l = 0$, that is, to the first term of
Eq.~(\ref{+-equat}). The partial $S$ matrix for the case of zero
angular momentum will be denoted by 
$S(E)\equiv S_{l=0}(E)$.

Because the $S$ matrix relates incoming and outgoing 
waves far outside the interaction region, it suffices to focus on 
the region $r>b$. In this region, we
have an incoming spherical wave $e^{-ikr}/r$ with 
amplitude ${\cal F}_2(k)$ and an outgoing spherical 
wave $e^{ikr}/r$ with amplitude ${\cal F}_1(k)$. Therefore, the 
$S$ matrix in the energy 
representation\cite{NEWTON,NUSSENZVEIG} is given by the ratio
\begin{equation}
	S(E)\equiv S(k)=-\frac{{\cal F}_1(k)}{{\cal F}_2(k)}
	\label{Sl0}\,,
\end{equation}
where $k$ is given by Eq.~(\ref{momentum}). It is easy to verify
that the $S$ matrix is unity when there is no potential,
\begin{equation}
	\lim_{V_0\to 0}S(k)=1 \, ,
\end{equation}
and that $S^*(k)S(k)=1$, that is, $|S(k)|=1$ for every $k$.

\begin{figure}[ht]
\hskip2cm\includegraphics[width=8.5cm]{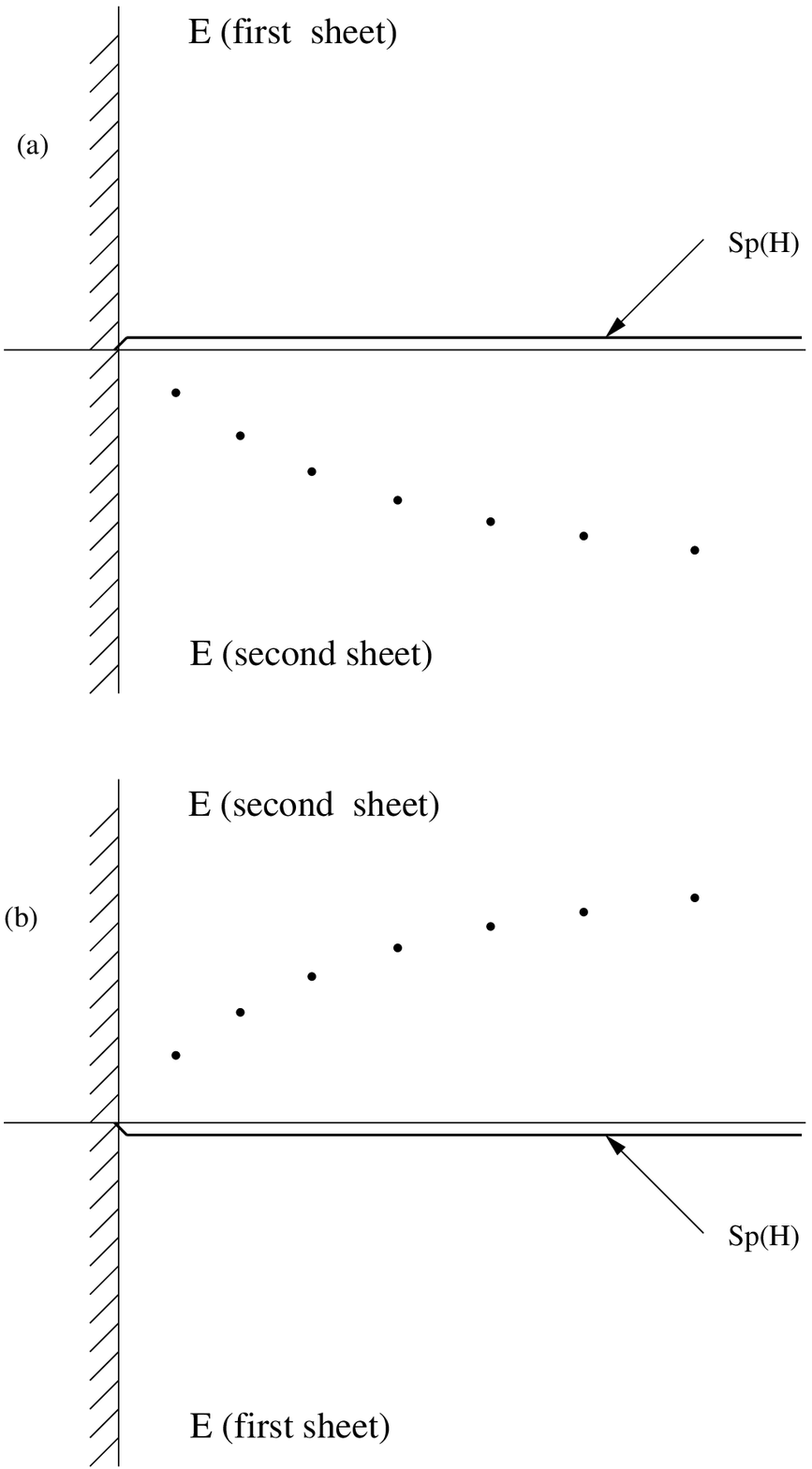}
\caption{The resonance energies of the square barrier potential.}
\label{epoles}
\end{figure}

The $S$ matrix $S(k)$ in Eq.~(\ref{Sl0}) is well defined for each
real value of $k$, because its denominator is never
zero when $k$ is real. The $l=0$ resonances are associated with the 
poles of the analytic continuation of $S(k)$ into the entire 
complex plane. The relation 
$k^2=2mE/\hbar^2$ provides a Riemann surface for this extension 
in a natural way. The analytic continuation of the 
numerator and the denominator of $S(k)$ yield two analytic functions 
${\cal F}_{1,2}(k)$. Therefore, the continuation of $S(k)$ 
is analytic except at its poles. These are precisely the 
zeros of the denominator of $S(k)$ (see Ref.~\onlinecite{MARSDEN}),
\begin{equation}
 {\cal F}_2(k)=0 \, ,
\end{equation}
which leads to
\begin{equation}
(1-\frac{Q}{k})e^{iQ(b-a)} [\sin (ka)+\frac{k}{iQ}\cos (ka)]
 + (1+\frac{Q}{k})e^{-iQ(b-a)} [\sin (ka)-\frac{k}{iQ}\cos (ka)
]=0 \, . \label{resonances}
\end{equation}
The solutions of Eq.~(\ref{resonances}) are the ($S$-matrix)
resonances of the square barrier potential.
Equation~(\ref{resonances}) has a denumerable infinite number of
complex resonance energy solutions. These solutions come in pairs
$E_{R}\pm i\Gamma_R /2$ (see Fig.~\ref{epoles}). The pole
$E_{R}-i\Gamma_R /2$ is associated with the decaying
part of the resonance, and is located on the lower half-plane
of the second sheet of the two-sheeted Riemann surface
corresponding to the square root mapping (see Fig.~\ref{epoles}a).
The pole
$E_{R}+i\Gamma_R /2$ is associated with the
growing part of the resonance, and is located on 
the upper half-plane of the second sheet of the Riemann surface (see 
Fig.~\ref{epoles}b). In the wave number plane, this pair of energy
poles corresponds to a pair of poles $\pm {\rm Re}(k)-i{\rm
Im}(k)$ in the lower half-plane that are mirror images of one
another with respect to the imaginary axis (see
Fig.~\ref{kpoles}). 

The width of the resonance increases as the energy increases, and therefore 
the lifetime $\tau =1/\Gamma_R$ decreases. The resonances whose
energies are below the threshold $E=V_0$ are close to the real
axis. As $E$ increases, the resonances move away from the
real axis toward infinity.

\subsection{Gamow Vector Approach}
\label{sec:gamow}
In this section, we determine the state vectors of the resonances
of the square barrier potential 
(\ref{potential}) as solutions of the time independent Schr\"odinger
equation with complex eigenvalues. The state vectors will be
eigenvectors as in Eq.~(\ref{tiSe}). Because in
Sec.~\ref{sec:time} we found all the eigenvectors of $H$
with the boundary conditions (\ref{boucodi}) and determined that
they all have positive real eigenvalues, we will have to change
these boundary conditions to purely outgoing boundary
conditions.\cite{KPS} We will see that there is a one-to-one
correspondence between the complex poles of the analytically
continued $S$ matrix and the complex eigenvalues obtained from
purely outgoing boundary conditions. These eigenvectors of the
Hamiltonian with complex eigenvalues, which are associated with the
poles of the $S$ matrix, are called Gamow vectors or Gamow kets.

The time independent Schr\"odinger equation for the Gamow vectors is given by 
\begin{equation}
	H|z_R\rangle =z_R|z_R\rangle \, , 
 \label{zg}
\end{equation}
where $z_R$ and $|z_R\rangle$ represent the complex resonance eigenvalue and 
the Gamow vector, respectively. In the position
representation, Eq.~(\ref{zg}) becomes
\begin{equation}
	\langle \vec{x}|H|z_R \rangle =
 z_R \langle \vec{x}|z_R \rangle,
 \label{tisefgv}
\end{equation}
or
\begin{equation}
	\left(-\frac{\hbar^2}{2m} \nabla^2 +V(\vec{x})\right)
	\langle \vec{x}|z_R\rangle=z_R\langle \vec{x}|z_R\rangle\,.
 \label{2.32}
\end{equation} 
After writing Eq.~(\ref{2.32}) in spherical coordinates and considering only 
the case of zero angular momentum, we obtain
\begin{equation}
	-\frac{\hbar^2}{2m}\frac{d^2}{dr^2}\chi (r;z_R)+V(r)\chi (r;z_R)
	=z_R\chi (r;z_R) \, .
	\label{Grse0}
\end{equation}
The solution of Eq.~(\ref{Grse0}) for each value of $z_R$ is the 
$s$-Gamow vector (more precisely, the radial part of the s-Gamow
vector in the position representation 
$\langle \vec{x}|z_R\rangle = \langle r,\theta, \phi |z_R,0,0\rangle 
=1/r \, \chi (r;z_R) Y_{0,0}(\theta, \phi)$). The solution 
of Eq.~(\ref{Grse0}) has the same form as Eq.~(\ref{rs})
\begin{equation}
 \chi (r;z_R)=\left\{ \begin{array}{ll}
 \alpha_1e^{ikr}+\beta_1e^{-ikr} &0<r<a \\
 \alpha_2e^{iQr}+\beta_2e^{-iQr} &a<r<b \\
 {\cal F}_1e^{ikr}+{\cal F}_2e^{-ikr} 
 &b<r<\infty \, , 
 \end{array} 
 \right. 
 \label{Grs}
\end{equation}
but now the wave number
\begin{equation}
	k=\sqrt{\frac{2m}{\hbar^2}z_R} 
 \label{cmomentum}
\end{equation}
of the resonance is {\it complex}, and 
\begin{equation}
	Q=\sqrt{k^2-\frac{2m}{\hbar^2}V_0}
 =\sqrt{\frac{2m}{\hbar^2}(z_R-V_0)} \, .
 \label{Qcmomentum}
\end{equation}

Because the second derivative of $\chi (r;z_R)$ must be well
defined, the solution (\ref{Grs}) must be continuous with
continuous derivatives as in (\ref{Dirboc2})--(\ref{pretlst}). 
For the Gamow vector we shall also choose the condition that the
radial part $\chi (r;z_R)$ vanishes at
the origin as in Eq.~(\ref{jaja}). However, at
infinity we choose the purely outgoing boundary condition, which
means that far from the potential region, the solution reduces
to an exponential of the type $e^{ikr}$, but not of the type 
$e^{-ikr}$. If $k$ is complex, this purely outgoing 
boundary condition does not mean that there are only 
outgoing waves. In fact, we have outgoing waves only when 
Re$(k)$ is positive, and incoming waves when Re$(k)$ is
negative. Thus the boundary conditions that we impose on the Gamow
vectors are
\begin{mathletters}
 \label{Gbc}
\begin{eqnarray}
	\chi (0;z_R)&=&0 \label{gvlov1} \\
	\chi(a-0;z_R)&=&\chi(a+0;z_R) \\
	\chi'(a-0;z_R)&=&\chi'(a+0;z_R) \\
	\chi(b-0;z_R)&=&\chi(b+0;z_R) \\
	\chi'(b-0;z_R)&=&\chi'(b+0;z_R) \\
	\chi(r;z_R)&\sim&e^{ikr} \, , \qquad \qquad r\to \infty \, .
\label{gvlov6} 
\end{eqnarray}
\end{mathletters}
The purely outgoing boundary condition (\ref{gvlov6}) is often
written as
\begin{equation}
 \lim_{r\to \infty} \frac{d\chi (r;z_R)}{dr} -ik\chi (r;z_R)=0 \, .
 \label{Mbc}
\end{equation} 
One can easily check that Eq.~(\ref{Mbc}) is equivalent to
Eq.~(\ref{gvlov6}).

The boundary condition (\ref{gvlov6}) leads by Eq.~(\ref{Grs}) to 
${\cal F}_2=0$. Because this condition is the same as the
condition (\ref{resonances}) for the complex poles of the
$S$ matrix (\ref{Sl0}), the set of complex eigenvalues $z_R$
must include the set of complex solutions of Eq.~(\ref{resonances})
(which are the $S$-matrix resonance poles). We shall show that
these two sets of solutions are the same.

If we define $\alpha =2i\alpha_1$, the boundary conditions (\ref{Gbc}) can 
be written in terms of the coefficients as
\begin{mathletters}
\begin{eqnarray}
	\alpha_2e^{iQa}+\beta_2e^{-iQa}&=&\alpha \sin (ka) \\
	iQ(\alpha_2e^{iQa}-\beta_2e^{-iQa})&=&\alpha k\cos (ka) \\
	{\cal F}_1e^{ikb}&=&\alpha_2e^{iQb}+\beta_2e^{-iQb} \\
	ik{\cal F}_1e^{ikb}&=&iQ(\alpha_2e^{iQb}-\beta_2e^{-iQb}) \, .
\end{eqnarray}
\end{mathletters} 
If we write this set of linear equations as a matrix equation, we
obtain
\begin{equation}
	\left(\begin{array}{cccc} \sin (ka)&0&-e^{iQa}&-e^{-iQa}\\
 k\cos (ka)&0&-iQe^{iQa}&iQe^{-iQa}\\
 0&e^{ikb}&-e^{iQb}&-e^{-iQb}\\
 0&ike^{ikb}&-iQe^{iQb}&iQe^{-iQb}
 \end{array}\right)
 \left(\begin{array}{cccc} \alpha \\ {\cal F}_1 \\\alpha_2 \\ \beta_2
 \end{array}\right)=
 \left(\begin{array}{cccc} 0\\ 0\\0 \\ 0
 \end{array}\right) \, .
\end{equation}
This is a homogeneous system of four equations with four unknowns.
The system has a non-trivial solution if, and only if, the determinant of the
coefficients is equal to zero,
\begin{equation}
	\left|\begin{array}{cccc} \sin (ka)&0&-e^{iQa}&-e^{-iQa}\\
 k\cos (ka)&0&-iQe^{iQa}&iQe^{-iQa}\\
 0&e^{ikb}&-e^{iQb}&-e^{-iQb}\\
 0&ike^{ikb}&-iQe^{iQb}&iQe^{-iQb}
 \end{array}\right|=0\,.
\end{equation}
Straightforward calculations lead to
\begin{equation}
 (1-\frac{Q}{k})e^{iQ(b-a)}\bigl[\sin (ka)+\frac{k}{iQ}\cos
(ka)\bigr]+
 (1+\frac{Q}{k})e^{-iQ(b-a)}\bigl[\sin (ka)-\frac{k}{iQ}\cos
(ka)\bigr]=0 \, .
 \label{gamowenrecom}
\end{equation}
Thus we obtain exactly the same condition as the condition
(\ref{resonances}) for the poles of the $S$ matrix. If we compare
the boundary conditions (\ref{boucodi}) imposed upon the Dirac
kets with the boundary conditions (\ref{Gbc}) imposed upon the
Gamow vectors, we can see that the boundary condition that singles
out the resonance spectrum is the purely outgoing boundary
condition (\ref{gvlov6}).

As we mentioned earlier, the solutions of Eq.~(\ref{gamowenrecom}) 
(or Eq.~(\ref{resonances})) always come in pairs. The eigenvector
corresponding to the complex eigenvalue $E_{R}-i\Gamma_R/2$ is the
decaying Gamow vector in the position representation, whose
radial part, up to a normalization factor, is 
\begin{equation}
	\chi^{\rm decaying}(r;z_R)=\left\{ \begin{array}{ll}
 \sin(k_{d}r) &0<r<a \\
 \alpha_2(k_{d})e^{iQ_{d}r}
 +\beta_2(k_{d})e^{-iQ_{d}r} &a<r<b \\
 {\cal F}_1(k_{d}) e^{ik_{d}r} &b<r<\infty \, , 
 \end{array} 
 \right.
	\label{dgv0p} 
\end{equation}
where $k_{d}=\sqrt{2m /\hbar^2 \, (E_{R}-i\Gamma_R /2)}$ and 
$Q_{d}^2=k_{d}^2-2m /\hbar^2 \, V_0$. 

For $E_{R}+i\Gamma_R /2$, we obtain the growing Gamow vector in
the position representation. Its radial part, up to a
normalization factor, is
\begin{equation}
	\chi^{\rm growing}(r;z_R^*)=\left\{ \begin{array}{ll}
 \sin(k_{g}r) &0<r<a \\
 \alpha_2(k_{g}) e^{iQ_{g}r}
 +\beta_2(k_{g}) e^{-iQ_{g}r} &a<r<b \\ 
 {\cal F}_1(k_g) e^{ik_{g}r} &b<r<\infty \, , 
 \end{array} 
 \right.
	\label{ggv0p} 
\end{equation}
where $k_{g}=\sqrt{2m/\hbar^2 \, (E_{R}+i\Gamma_R /2)}$ and 
$Q_{g}^2=k_{g}^2-2m/ \hbar^2 \, V_0$. 

[Gamow eigenfunctions are not square integrable, that is,
they do not belong to the Hilbert space. Thus, as Dirac kets, they must
be handled by methods that extend those available in the Hilbert space 
framework. The Rigged Hilbert Space provides those methods
(see for example, Refs.~\onlinecite{DIS} and \onlinecite{BG}).]

\subsection{Green Function Approach}
\label{sec:green}
In this section, we perform the calculations for the resonance
energies using the Green function method. The radial Green
function satisfies (see for
instance, Refs.~\onlinecite{NEWTON} and \onlinecite{CSF}) 
\begin{equation}
 \left( -\frac{\hbar^2}{2m}\frac{\partial^2}{\partial r^2}+V(r)-E\right)
 G(r,r';E)=-\delta (r-r'),
 \label{greene}
\end{equation}
subject to various boundary conditions. Equation~(\ref{greene}) says
that for $r\neq r'$, $G(r,r';E)$ obeys the radial Schr\"odinger 
equation~(\ref{rSe0}). At $r=r'$ it is continuous, but its derivative 
has a discontinuity due to the delta function,
\begin{equation}
	\frac{\partial}{\partial r}G(r'+0,r';E)-
	\frac{\partial}{\partial r}G(r'-0,r';E)=\frac{2m}{\hbar^2} \, .
\end{equation}
We want $G$ to satisfy the same kind of
boundary conditions at zero and at infinity that we imposed on the 
other methods. We shall also require $G$ to be regular at zero and
a purely outgoing boundary condition at infinity. The expression
for this Green function is\cite{NEWTON,CSF}
\begin{equation}
	G(r,r';E)=\frac{2m}{\hbar^2}
 \frac{\chi (r_<;E) \psi(r_>;E)}{W(\chi ,\psi )} \, ,
 \label{G+l=so}
\end{equation}
where $r_<$ refers to the minimum and $r_>$ to the maximum of $r$
and
$r'$. The function $\chi$ is the solution of Eq.~(\ref{rSe0}) that
vanishes at the origin, 
$\psi$ is the solution of Eq.~(\ref{rSe0})
that satisfies a purely outgoing boundary condition at infinity,
and
$W(\chi,\psi)$ is their Wronskian. In this case $\chi$ is just the 
solution in Eq.~(\ref{sss0}). The function $\psi$
satisfies Eq.~(\ref{rSe0}) and the boundary conditions
\begin{mathletters}
 \label{bbb2}
\begin{eqnarray}
	\psi (a-0;E)&=&\psi (a+0;E) \label{psibc1} \\
	\psi '(a-0;E)&=&\psi '(a+0;E) \\
	\psi (b-0;E)&=&\psi (b+0;E) \\
	\psi '(b-0;E)&=&\psi '(b+0;E) \label{psibc4} \\
	\psi (r;E)&\sim&e^{ikr} \, , \qquad \quad r\to \infty \, ,
 \label{psibc5} 
\end{eqnarray}
\end{mathletters} 
The solution $\psi$ is then given by
\begin{equation}
	\psi (r;k)=\left\{ \begin{array}{ll}
 a_1(k)e^{ikr}+b_1(k)e^{-ikr} &0<r<a \\
 a_2(k)e^{iQr}
 +b_2(k)e^{-iQr} &a<r<b \\
 e^{ikr}  &b<r<\infty \, ,
 \end{array} 
 \right. \label{psifun}
\end{equation}
where the coefficients $a(k)$ and $b(k)$ are functions of $k$
that make 
$\psi$ satisfy the boundary conditions of Eq.~(\ref{bbb2}). The
Wronskian of 
$\chi$ and $\psi$ is
\begin{equation}
	W(\chi ,\psi )(k)= \chi (r)\psi '(r)-\chi '(r)\psi (r)=
 2ik {\cal F}_2(k)\,.
 \label{Wronsol=0}
\end{equation}
From Eqs.~(\ref{G+l=so}) and (\ref{Wronsol=0}), we obtain the Green function,
\begin{equation}
	G(r,r';k)=\frac{2m}{\hbar^2}
 \frac{\chi (r_<;k) \psi(r_>;k)}{2ik{\cal F}_2(k)} \, .
 \label{GFUNSO}
\end{equation}
The poles of this Green function are simply the zeros of its 
denominator.\cite{MARSDEN} Those poles are the same as the
$S$-matrix poles (\ref{resonances}) and as the complex eigenvalues
$z_R$ in Eq.~(\ref{gamowenrecom}). 

It is worth noting that the Green function (\ref{GFUNSO}) is
an outgoing Green function only when ${\rm Re} (k)$ is positive. If
${\rm Re}(k)$ is negative, the Green function (\ref{GFUNSO}) has a
purely incoming character (cf.~Sec.~\ref{sec:PAGV}). Also, as $k$
approaches the real positive
$k$ axis or as $E$ approaches the right-hand cut from above, the
Green function $G$ becomes the {\it outgoing} Green function $G^+$;
as $k$ approaches the negative real axis (from above) or $E$ the
right-hand cut from below, $G$ becomes the {\it incoming} Green
function 
$G^-$.\cite{NEWTON}

The residue of the Green function (\ref{GFUNSO}) at a decaying
resonance wave number is given by
\begin{equation}
	\mbox{res} \left[ G(r,r';k) \right]_{k=k_d}=
 \frac{2m}{\hbar^2}\frac{1}{2ik_d{\cal F}_1(k_d){\cal F}_2'(k_d)}
 \chi^{\rm decaying}(r_<;k_d) \chi^{\rm decaying}(r_>;k_d) \, .
\end{equation}
A similar relation between the residue of $G(r,r';k)$ at the
growing wave number pole $k_g$ and the growing Gamow
eigenfunction (\ref{ggv0p}) holds.

\subsection{Complex Basis Vector Expansion}
\label{sec:complex}
The Dirac kets are basis vectors that were used to expand a wave 
function $\varphi$ as in Eq.~(\ref{vep0}). This expansion is 
known as the Dirac basis vector expansion.\cite{DIRAC} The Gamow
vectors are also basis vectors. The expansion generated by the
Gamow vectors is called the complex basis vector expansion.
However, the Gamow vectors do not form a complete basis. The
complex basis vector expansion needs an additional set of Dirac
kets corresponding to the energies that lie in the negative real
axis of the second sheet of the Riemann surface.\cite{BOHM,DIS} This
has been realized by other 
authors\cite{MONDRAGON,GASTON,BERGGREN} who used the Green
function approach. In this section, we shall expand a wave
function in terms of the Gamow vectors (which contain the
resonance contribution) and a continuous set of Dirac kets (which
is interpreted as a background contribution). 

[Rigorously speaking, the complex basis vector expansion is not
valid for every normalizable wave function, that is, for every
element of the Hilbert space, but only for those square integrable
functions that are also Hardy functions.\cite{DIS,BG,GADELLA}]

\begin{figure}[ht]
\hskip2cm\includegraphics[width=8.5cm]{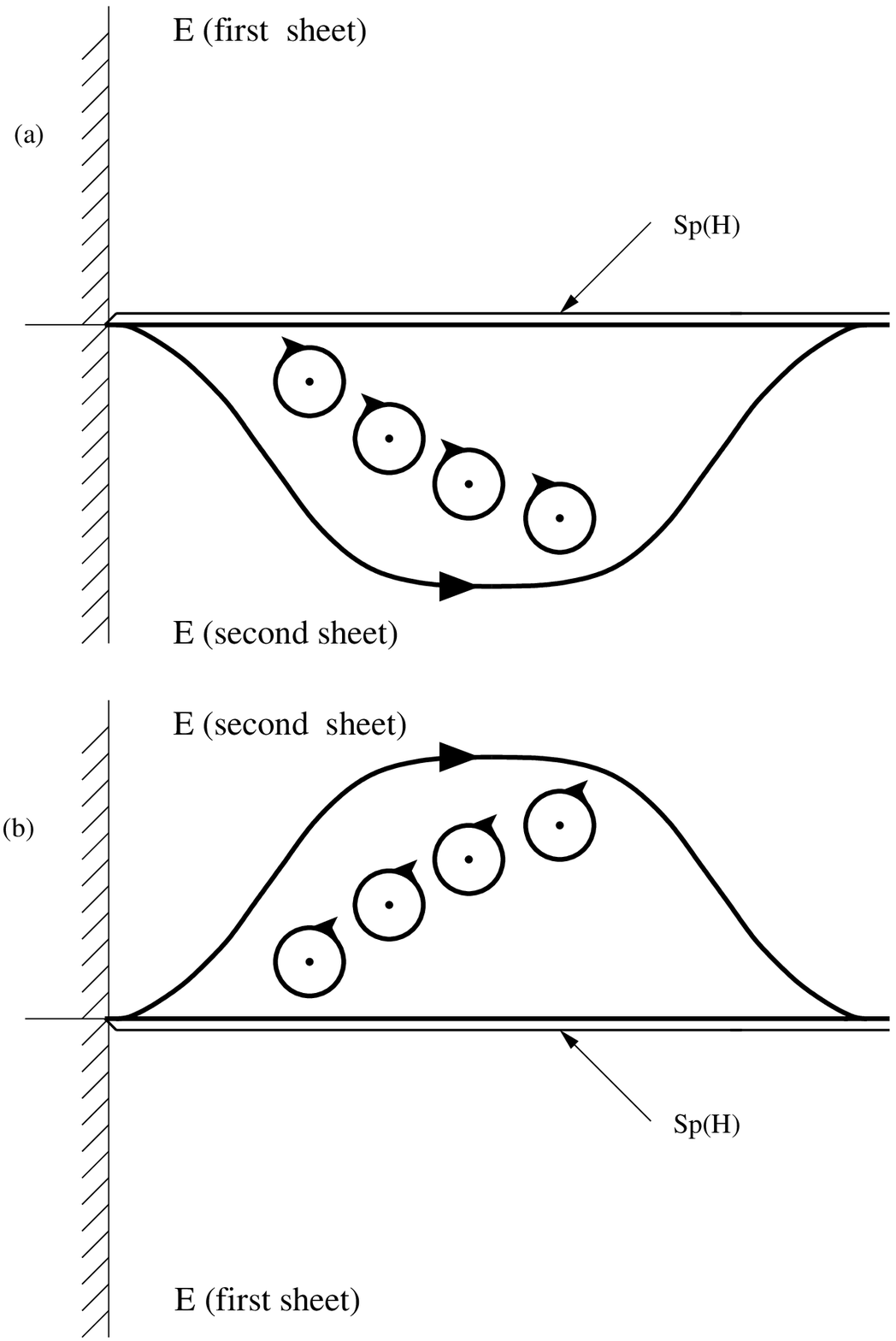}
\caption{Deformation of the path of integration into the second sheet of the 
energy Riemann surface; (a) for the decaying states and 
(b) for the growing states.}
\label{contour}
\end{figure}

Experimentally, we can only measure the probability for an event to take 
place. In a scattering experiment this is the transition
probability from an in-state $\varphi^+$ to an out-state $\psi^-$.
The amplitude of this probability is calculated from the scalar
product between 
$\varphi^+$ and $\psi^-$. In terms of the quantities that 
we already have calculated, this amplitude is given
by\cite{BOHM,DIS} 
\begin{equation}
	\left( \psi^-,\varphi^+\right)=\!
	\int_0^{\infty} \! \langle \psi^-|E^-\rangle S(E)
	\langle E^+|\varphi^+\rangle\, dE \, .
	\label{superequation}
\end{equation}
We now extract the resonance contribution from
Eq.~(\ref{superequation}). This resonance contribution is carried by
the Gamow vectors. In order to do so, we deform the contour of
integration into the lower half-plane of the second sheet of the
Riemann surface for the $S$ matrix, where the resonances poles are
located (see Fig.~\ref{contour}a). If we use the results in
Ref.~\onlinecite{GADELLA}, we can write Eq.~(\ref{superequation}) as
\begin{equation}
 \left( \psi^-,\varphi^+\right)
 =\!\int_0^{-\infty}\! \langle \psi^-|E^+\rangle 
 \langle E^+|\varphi^+\rangle dE 
 -2\pi i \sum_{n=0}^{\infty}r_n
 \langle \psi^-|z_{d,n}^-\rangle\langle^+z_{d,n}|\varphi^+\rangle ,
 \label{zigzag}
\end{equation}
where $z_{d,n}=E_{n}-i\Gamma_n/2$ denotes the $n$th decaying
pole and 
$r_n$ denotes the residue of $S(E)$ at $z_{d,n}$. In
Eq.~(\ref{zigzag}), the integration is done on the negative real
semiaxis of the second sheet of the Riemann surface. The series 
in Eq.~(\ref{zigzag}) can be shown to be convergent.\cite{GADELLA}
Omitting
$\psi^-$ in~(\ref{zigzag}), we obtain the complex basis vector
expansion for the in-states, 
\begin{equation}
	\varphi^+
	=\!\int_0^{-\infty} \! |E^+\rangle 
	\langle^+E|\varphi^+\rangle dE 
	-2\pi i\sum_{n=0}^{\infty}r_n
	|z_{d,n}^- \rangle\langle^+z_{d,n}|\varphi^+\rangle \, . 
 \label{states}
\end{equation}
In Eq.~(\ref{states}), the infinite sum contains the resonances
contribution, while the integral is interpreted as the background
contribution.

Similarly, we obtain the complex basis vector expansion for the
out-state 
$\psi^-$,\cite{GADELLA} but now we deform the contour of
integration into the upper half-plane of the second sheet of the
Riemann surface, where the growing resonance poles are located (see
Fig.~\ref{contour}b). The result is
\begin{equation}
	\psi^-
	=\! \int_0^{-\infty} \!|E^-\rangle \langle E^-|\psi^- \rangle dE 
	+2\pi i\sum_{n=0}^{\infty}r_n^*
	|z_{g,n}^*{^+}\rangle\langle^-z^*_{g,n}|\psi^-\rangle \, ,
 \label{observables}
\end{equation}
where $z_{g,n}^* =E_{n}+i\Gamma_n/2$ is the $n$th growing pole,
and 
$r_n^*$ is the residue of $S(E)$ at $z_{g,n}^*$. The integration in
Eq.~(\ref{observables}) is performed on the negative real semiaxis
of the second sheet of the Riemann surface. The series in
Eq.~(\ref{observables}) has been shown to be
convergent.\cite{GADELLA}

\section{Time Asymmetry of the Purely Outgoing Boundary Condition}
\label{sec:PAGV}
The semigroup time evolution of the Gamow vectors expresses the
time asymmetry built into them.\cite{BOHM,DIS,BG} We will show
here that the purely outgoing boundary condition that singles out
the resonance energies also has an intrinsic time asymmetry. To be more 
precise, we will show that the purely outgoing
boundary condition should be read as {\it purely outgoing} only for
the decaying part of a resonance and as {\it purely incoming} for
the growing part of the resonance. Obviously, the purely incoming
boundary condition is the time reversed of the purely outgoing one. Therefore
the growing Gamow vector can be viewed as the time reversed of the decaying 
Gamow vector.

\subsection{Outgoing Boundary Condition in Phase}
\label{sec:outgoing}
\begin{figure}[ht]
\hskip3cm\includegraphics[width=8.5cm]{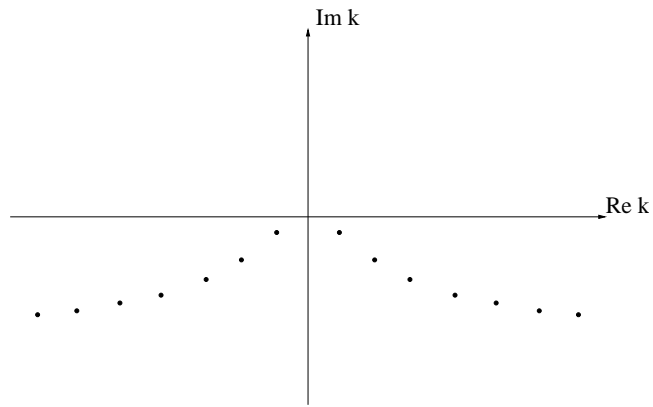}
\caption{The resonance wave numbers of the square barrier potential.}
\label{kpoles}
\end{figure}

First, we study the meaning of the purely outgoing boundary condition when
it is imposed on the decaying part of the resonance. The complex
energy associated with the decaying part of a resonance is
$z_d=E_R-i\Gamma_R/2$ ($E_R, \Gamma_R>0$), which lies in the lower
half-plane of the second sheet of the Riemann surface (see
Fig.~\ref{epoles}a). Its wave number 
$k_d={\rm Re}(k)-i{\rm Im}(k)$ (Re$(k)$, Im$(k)>0$) lies in the fourth quadrant
of the wave number plane (see Fig.~\ref{kpoles}). The decaying
Gamow vector $\chi^{\rm decaying}$ in Eq.~(\ref{dgv0p}) was obtained
after imposing the purely outgoing boundary condition
(\ref{gvlov6}) on Eq.~(\ref{Grs}). If we had not imposed this
condition, we would had obtained a solution of the form 
(\ref{Grs}), and every complex number would had been an eigenvalue
of the Hamiltonian. In the region $r>b$, this solution would had 
been the sum of two linearly independent solutions:
\begin{eqnarray}
 \chi_{\rm incoming}^{\rm decaying}(r,t)&=&
 {\cal F}_2e^{-ik_dr}e^{-iz_dt/\hbar} \nonumber \\
 &=&\left( {\cal F}_2e^{-{\rm Im}(k)r-\Gamma_R t/(2\hbar)}\right) 
 e^{-i{\rm Re}(k)r-iE_Rt/\hbar} \, ,
 \qquad r>b \, ,
 \label{di}
\end{eqnarray}
which we call the incoming decaying Gamow vector, and
\begin{eqnarray}
 \chi_{\rm outgoing}^{\rm decaying}(r,t)&=&{\cal F}_1e^{ik_dr}e^{-iz_dt/\hbar} 
 \nonumber \\
 &=&\left( {\cal F}_1 e^{{\rm Im}(k)r-\Gamma_R t/(2\hbar)}\right) 
 e^{i{\rm Re}(k)r-iE_Rt/\hbar} \, , \qquad r>b \, , 
 \label{do}
\end{eqnarray}
which we call the outgoing decaying Gamow vector. These names come
from the standard interpretation of plane waves with a complex
exponent (see for instance
Ref.~\onlinecite{BORN}): the exponential with a purely imaginary exponent (the
term that carries the phase) is interpreted as the term that
governs the propagation of the wave, and the exponential with the
real exponent is interpreted as the term that just changes the
amplitude of the wave on the surfaces of equal phase.\cite{BORN}
We are going to interpret Eqs.~(\ref{di}) and (\ref{do}) in the
same fashion. The terms between parentheses in Eqs.~(\ref{di}) and
(\ref{do}) determine the amplitude of the waves. The propagation of
$\chi_{\rm outgoing}^{\rm decaying}$ is governed by
$e^{i{\rm Re}(k) r}e^{-iE_{R}t/\hbar}$, and therefore 
$\chi_{\rm outgoing}^{\rm decaying}$ is an outgoing wave (in
phase). Analogously, the propagation of $\chi_{\rm incoming}^{\rm
decaying}$ is governed by $e^{-i{\rm Re}(k) r}e^{-iE_{R}t/\hbar}$,
and thus 
$\chi_{\rm incoming}^{\rm decaying}$ is an incoming wave (in
phase). Imposing the purely outgoing boundary condition ${\cal
F}_2 =0$ is tantamount to forbidding $\chi_{\rm incoming}^{\rm
decaying}$. Thus for the decaying part of the resonance, the
purely outgoing boundary condition allows only purely outgoing
waves.

The meaning of the purely outgoing boundary condition applied to
the growing part of the resonance is the opposite. The
growing energy eigenvalue $z_g=E_R+i\Gamma_R/2$ ($E_R$,
$\Gamma_R>0$) lies in the upper half-plane of the second sheet of
the Riemann surface (see Fig.~\ref{epoles}b), and its wave number
$k_g=-{\rm Re}(k)-i{\rm Im}(k)$ (Re$(k)$, Im$(k)>0$) lies in the
third quadrant of the wave number plane (see Fig.~\ref{kpoles}). The
growing Gamow vector $\chi^{\rm growing}$ in Eq.~(\ref{ggv0p}) was
obtained after imposing the condition ${\cal F}_2=0$ on 
Eq.~(\ref{Grs}). If we had not imposed this condition, in the
region
$r>b$, the solution would had been the sum of two linearly 
independent solutions:
\begin{eqnarray}
 \chi_{\rm incoming}^{\rm growing}(r,t)&=&
 {\cal F}_1e^{ik_gr}e^{-iz_gt/\hbar} 
 \nonumber \\
 &=&\left( {\cal F}_1 e^{{\rm Im}(k)r+\Gamma_R t/(2\hbar)}\right) 
 e^{-i{\rm Re}(k)r-iE_Rt/\hbar} \, , \qquad r>b \, ,
 \label{gi}
\end{eqnarray}
which we call the incoming growing Gamow vector, and
\begin{eqnarray}
 \chi_{\rm outgoing}^{\rm growing}(r,t)&=&{\cal F}_2 e^{-ik_gr}e^{-iz_gt/\hbar} 
 \nonumber \\
 &=&\left( {\cal F}_2 e^{-{\rm Im}(k)r+\Gamma_R t/(2\hbar)}\right) 
 e^{i{\rm Re}(k)r-iE_Rt/\hbar}\, , \qquad r>b \, , 
 \label{go}
\end{eqnarray}
which we call the outgoing growing Gamow vector. The names also come
from the standard interpretation\cite{BORN} of plane waves with 
a complex exponent. Therefore, the purely outgoing boundary
condition 
${\cal F}_2 =0$, when applied to the growing part of a resonance,
bans $\chi_{\rm outgoing}^{\rm growing}$ and allows only purely
incoming waves.

\subsection{Outgoing Boundary Condition in Probability Density}
\label{Ovouiampi}
In Sec.~\ref{sec:outgoing}, we showed how the time asymmetry built into the 
purely outgoing boundary condition affected the phase of the Gamow 
vectors. In this section, we show the same time asymmetry but now 
consider the probability density of the Gamow vectors.

For the decaying part of the resonance, the probability
densities (before imposing the purely outgoing boundary condition)
are obtained by taking the square of the absolute value of
Eq.~(\ref{di})
\begin{eqnarray}
 \rho_{\rm incoming}^{\rm decaying}(r,t)=
 |\chi_{\rm incoming}^{\rm decaying}(r,t)|^2&=& 
 |{\cal F}_2|^2 e^{-2{\rm Im}(k)r-\Gamma_R t/\hbar} \nonumber \\ 
 &=& |{\cal F}_2|^2 e^{-\Gamma_R /\hbar (t+r/v)}\, , \qquad r>b \, , 
 \label{dia}
\end{eqnarray}
which we call the incoming decaying probability density. Similarly,
the square of the absolute value of Eq.~(\ref{do}) yields
\begin{eqnarray}
 \rho_{\rm outgoing}^{\rm decaying}(r,t)=|\chi_{\rm outgoing}^{\rm decaying}(r,t)|^2
 &=& |{\cal F}_1|^2 e^{2{\rm Im}(k)r-\Gamma_R t/\hbar} \nonumber \\
 &=& |{\cal F}_1|^2 e^{-\Gamma_R / \hbar (t-r/v)}\, , \qquad r>b \, , 
 \label{doa}
\end{eqnarray}
which we call the outgoing decaying probability density
($v=\Gamma_R /(2\hbar {\rm Im}(k))$). By imposing the purely outgoing
boundary condition ${\cal F}_2=0$, we allow only Eq.~(\ref{doa})
and forbid Eq.~(\ref{dia}), which we interpret by
saying that we have a purely outgoing probability density
condition for the decaying part of the resonance. 

For the growing part of the resonance, the probability
densities (before imposing ${\cal F}_2=0$) are the square of
the absolute value of Eq.~(\ref{gi}),
\begin{eqnarray}
 \rho_{\rm incoming}^{\rm growing}(r,t)=|\chi_{\rm incoming}^{\rm growing}(r,t)|^2 
 &=& |{\cal F}_1|^2 e^{2{\rm Im}(k)r+\Gamma_R t/\hbar} \nonumber \\
 &=& |{\cal F}_1|^2 e^{\Gamma_R /\hbar (t + r/v )} \, , \qquad r>b \, ,
 \label{gia}
\end{eqnarray}
which we call the incoming growing probability density. Similarly,
from Eq.~(\ref{go}), we have
\begin{eqnarray}
 \rho_{\rm outgoing}^{\rm growing}(r,t)=|\chi_{\rm outgoing}^{\rm growing}(r,t)|^2&=&
 |{\cal F}_2|^2 e^{-2{\rm Im}(k)r+\Gamma_R t/\hbar} \nonumber \\
 &=&|{\cal F}_2|^2 e^{\Gamma_R /\hbar (t-r/v)} \, , \qquad r>b \, , 
 \label{goa}
\end{eqnarray}
which we call the outgoing growing probability density. For this 
growing part, the condition ${\cal F}_2=0$ only allows waves
with purely incoming probability densities.

In short, the purely outgoing boundary condition (\ref{gvlov6})
must be read as purely outgoing (in phase or in probability
density) only for the decaying part of the resonance and as
purely incoming (in phase or in probability density) for the
growing part of the resonance. 

\subsection{Exponential Decay Law of the Gamow Vectors}
\label{Compuprova}
We want to determine the probability ${\cal P}_{\Delta r_0}(t)$ of
detecting the decaying state within a shell of width $\Delta r_0$
outside the potential region ($r>b$). This is the probability that
is measured by the counting rate of a detector placed, for
example, outside a radioactive nucleus from which an
$\alpha$-particle is emitted. We assume that the detector surrounds
the nucleus completely and that is at a distance 
$r_0>b$ from the center $r=0$.

Theoretically, the probability ${\cal P}_{\Delta r_0}(t)$ to
observe an in-state $\varphi^+$ at time $t$ within the interval
$\Delta r_0$ around the surface $r=r_0$ is given by
\begin{equation}
 {\cal P}_{\Delta r_0}(t)=\!
 \int \!d\Omega \!\int_{\Delta r_0}\! r^2 \,  dr
 |\langle r,\theta ,\phi |\varphi^+(t) \rangle |^2 \, .
 \label{pinpsi-}
\end{equation}
Experimentally, the probability of finding the decaying state 
particle around $r_0$, that is, the counting rate of the detector,
is not defined for all times $t$: a resonance must be first
prepared before the system can decay. The time at which the
preparation of the resonance is finished and at which the decay
starts can be chosen arbitrarily (we choose it to be 
$0$). For example, the 
$\alpha$ particle emitted by an $\alpha$-unstable nucleus travels at speed 
$v=\Gamma_R /(2\hbar {\rm Im}(k))$ and reaches the point $r_0$ at
the time 
$t(r_0)=r_0/v$. For times less than $t(r_0)$, the $\alpha$ particle
is not there yet, and therefore the counting rate measured by a
detector placed at $r_0$ is zero for times $t<r_0/v$. Whatever
would have been counted by the detector before the instant
$t(r_0)$ at $r_0$ cannot be connected with the decaying state.
Thus the theoretical probability to detect a resonance at $r_0$
should be zero for $t<r_0/v$. This is an instance of the time 
asymmetry built into a decaying process.

Experimentally as well, the decay of unstable systems usually follows the
exponential law
(cf.~Refs.~\onlinecite{ROLLEFSON,SUPON,DESMARAIS,RUDDICK}).

[Hilbert space cannot accommodate either the time asymmetry of 
${\cal P}_{\Delta r_0}(t)$\cite{BGM} or the exponential decay 
law.\cite{KHALFIN} To account for these two features, we
should use the Rigged
Hilbert Space.\cite{BOHM,DIS,BG} In this formulation, the
Gamow vectors have an asymmetric time evolution given by a
semigroup, which accounts for the time asymmetry of a resonant
process. The behavior of the semigroup evolution is in contrast to
the time-symmetric Hilbert space time evolution, which is given by
a group.]

We are going to show that the exponential decay law holds if we consider only 
the resonance (Gamow vector) contribution to the probability 
(\ref{pinpsi-}). In Sec.~\ref{sec:complex}, we used the Gamow
vectors as basis vectors to expand the in-state $\varphi^+$ in
terms of the background and the resonance contribution (see
Eq.~(\ref{states})). To calculate the resonance
contribution to the probability (\ref{pinpsi-}), we approximate
$\varphi^+$ by the Gamow vector by neglecting the background term
in Eq.~(\ref{states}),
\begin{equation}
 \varphi^+(r,\theta ,\phi ) \simeq \psi^D(r,\theta ,\phi )=
 \frac{\chi^{\rm decaying}(r)}{r}Y_{0,0}(\theta ,\phi ) \, .
\end{equation}
Thus the resonance contribution to the probability is
\begin{equation}
 {\cal P}_{\Delta r_0}(t)\simeq \!\int \! d\Omega \! \int_{\Delta
r_0} \! r^2 dr
 |\langle r,\theta,\phi |\psi^D(t)\rangle |^2 \, . 
 \label{probability}
\end{equation}
The time evolution of the Gamow vector\cite{BOHM,DIS,BG}
is given by 
\begin{equation}
 \psi^D(t)=e^{-iHt/\hbar}\psi^D=
 e^{-i(E_Rt-i\Gamma_R /2)t/\hbar }\psi^D \, ,
 \label{semigrou}
\end{equation}
and therefore
\begin{equation}
 \langle r,\theta, \phi |\psi^D(t)\rangle =
 e^{-i(E_R-i\Gamma_R /2) t/\hbar}\,
 \frac{\chi^{\rm decaying}(r)}{r}Y_{0,0}(\theta, \phi ) \label{tGv} \, .
\end{equation} 
If we substitute Eq.~(\ref{tGv}) into Eq.~(\ref{probability}), we
obtain
\begin{eqnarray}
 {\cal P}_{\Delta r_0}(t)&\simeq & 
 |e^{-\Gamma_R /(2\hbar) t}|^2 \! \int_{\Delta r_0} \!
 dr |\chi^{\rm decaying}(r)|^2 \nonumber \\
 &=&e^{-\Gamma_R t/\hbar}\!\int_{\Delta r_0} \!dr |{\cal F}_1(k)|^2 
 |e^{i({\rm Re}(k)-i{\rm Im}(k))r}|^2 \nonumber \\
 &=&e^{-\Gamma_R t/\hbar}|{\cal F}_1(k)|^2 \!
 \int_{r_0}^{r_0 +\Delta r_0}\! dr e^{2{\rm Im}(k)r} \nonumber \\
 &=&e^{-\Gamma_R t/\hbar}|{\cal F}_1(k)|^2 e^{2{\rm Im}(k)r_0} \,
 \frac{e^{2{\rm Im}(k)\Delta r_0}-1}{2{\rm Im}(k)} \nonumber \\
 &\simeq & e^{-\Gamma_R t/\hbar}|{\cal F}_1(k)|^2 e^{2{\rm Im}(k)r_0} 
 \, 
 \Delta r_0 \nonumber \\
 &=&|{\cal F}_1(k)|^2 \Delta r_0 \, e^{-\Gamma_R/\hbar(t-r_0/v)}\, , 
 \qquad t>r_0/v \, , 
\end{eqnarray}
where we have used the approximation $\Delta r_0$ small in the next to the
last step. Therefore,
\begin{equation}
 {\cal P}_{\Delta r_0}(t)\simeq
 |{\cal F}_1(k)|^2\Delta r_0 e^{-\Gamma_R/\hbar(t-r_0/v)}\, , 
 \qquad t>r_0/v \, .
 \label{probabitgvr.v}
\end{equation}
Equation~(\ref{probabitgvr.v}) represents the resonance contribution
to the counting rate measured by a detector placed at $r_0$. This
resonance contribution reaches its maximum at $t=r_0/v$ and
decreases exponentially as time goes on. Therefore, the Gamow
vector (resonance) contribution to the probability ${\cal
P}_{\Delta r_0}(t)$ follows the exponential decay law.

\section{Conclusions}
We have studied the different kinds of boundary
conditions that we need to impose on the time independent
Schr\"odinger equation in order to obtain either Dirac kets
(scattering states) and the physical spectrum or Gamow vectors
(resonance states) and the resonance spectrum. By
imposing the boundary condition of boundedness (\ref{bc}), we obtain
the Dirac kets and the spectrum $[0,\infty)$. And by
imposing the purely outgoing boundary condition (\ref{gvlov6}), we
obtain the Gamow vectors and the resonance spectrum of
Fig.~\ref{epoles}. This purely outgoing boundary condition
produces the same resonance spectrum as the
$S$ matrix and the Green function. 

The Gamow vectors have been used as basis vectors in the complex basis vector
expansions (\ref{states}) and (\ref{observables}). However, they do not form 
a complete basis, and therefore a continuous set of Dirac kets was added to 
complete them. The expansions (\ref{states}) and (\ref{observables})
extract the resonance contribution out of the normalized in- 
and out-states, respectively.

We have uncovered the time asymmetry that arises from the purely outgoing 
boundary condition. We have seen that the purely outgoing boundary condition
should be read as {\it purely outgoing} only for the decaying
part of the resonance, and as {\it purely incoming} for the
growing part of the resonance.

The exponential law has been shown to hold if the background term
of the complex basis vector expansion is neglected---only the
resonance (Gamow vector) contribution to the probability is taken
into consideration. These conclusions are not restricted to the
example given here and are in fact applicable to a large class of
potentials that includes potentials with compact support.

\section*{Acknowledgments}
The authors wish to thank A.~Bohm, N.~Harshman, H.~Kaldass, C.~Koeninger,
M.~Mithaiwala, C.~Puntmann, B.~A.~Tay, and S.~Wickramasekara for
proofreading the paper and making invaluable suggestions. One of
the authors (RM) wishes to thank the Physics
Department of the University of Texas at Austin for their
hospitality during the elaboration of part of this work. Special thanks to
Prof.~H.~Gould for his numerous style suggestions, which have made the paper 
much more readable.

Financial support from the E.U.~under TMR contract number
ERBFMRX-CT96-0087 The Physics of Quantum Information, from the
Welch Foundation, from DGICYT PB98-0370, from DGICYT PB98-0360,
and from La Junta de Castilla y Le\'on Project PC02 1999 is
gratefully acknowledged.

\appendix
\section{Higher Order Angular Momentum}
\label{sec:extension}
We extend the analysis done in Sec.~\ref{sec:resonances}
to the case in which the angular momentum $l>0$. 

\subsection{Eigenkets}
\label{eigenkets}
The radial part $\chi_l(r;E)$ of $\langle r,\theta , \phi |E,l,m \rangle$
satisfies Eq.~(\ref{baba}). The solution of this equation can be
written as
\begin{equation}
 \chi_l(r;E)=\left\{ \begin{array}{ll} 
 \alpha_{l1} \hat{h}_l^+(kr)+\beta_{l1} \hat{h}_l^-(kr) & 0<r<a \\
	\alpha_{l2} \hat{h}_l^+(Qr)+\beta_{l2} \hat{h}_l^-(Qr) & a<r<b \\
	{\cal F}_{l1} \hat{h}_l^+(kr)+{\cal F}_{l2} \hat{h}_l^-(kr) 
 & b<r<\infty \, ,
 \end{array} \right.
 \label{Rlson}
\end{equation}
where $k$ and $Q$ are given by Eqs.~(\ref{momentum}) and
(\ref{Qmomentum}) respectively. The functions $\hat{h}_l^+(kr)$
and $\hat{h}_l^-(kr)$ are the Riccati-Hankel
functions.\cite{WATSON} The boundary conditions that $\chi_l(r;E)$
is to satisfy are given in Eq.~(\ref{boucodi}). If we write 
$\alpha_l =2i\alpha_{l1}$, these 
boundary conditions can be written in terms of the coefficients as
\begin{mathletters}
\begin{eqnarray}
	 \alpha_{l2} \hat{h}_l^+(Qa)+\beta_{l2} \hat{h}_l^-(Qa)&=&
 \alpha_l \hat{j}_l(ka) \\
	\alpha_{l2}\hat{h}_l^{+\prime}(Qa)+
 \beta_{l2}\hat{h}_l^{-\prime}(Qa)&=&
 \frac{k}{Q}\alpha_l \hat{j}_l^{\prime}(ka) \\
	{\cal F}_{l1} \hat{h}_l^+(kb)+{\cal F}_{l2} \hat{h}_l^-(kb)&=&
	\alpha_{l2}\hat{h}_l^+(Qb)+\beta_{l2} \hat{h}_l^-(Qb) \\
	{\cal F}_{l1}\hat{h}_l^{+\prime}(kb)+
 {\cal F}_{l2}\hat{h}_l^{-\prime}(kb)&=&
	\frac{Q}{k}(\alpha_{l2}\hat{h}_l^{+\prime}(Qb)+
 \beta_{l2}\hat{h}_l^{-\prime}(Qb)) \, , 
\end{eqnarray}
\end{mathletters} 
where $\hat{j}_l=\frac{1}{2i}(\hat{h}_l^+ - \hat{h}_l^-)$,
$\hat{h}_l^{+\prime}(kb)=\frac{d\hat{h}_l^{+}(z)}{dz}|_{z=kb}$, and so 
on. Then 
\begin{mathletters}
\begin{eqnarray}
\alpha_{l2}(k)&=&\alpha_l(k)
	\frac{\hat{j}_l(ka)\hat{h}_l^{-\prime}(Qa)-
 \frac{k}{Q}\hat{j}_l^{\prime}(ka)\hat{h}_l^-(Qa)}
	{\hat{h}_l^+(Qa)\hat{h}_l^{-\prime}(Qa)-
 \hat{h}_l^{+\prime}(Qa)\hat{h}_l^-(Qa)} 
	\label{coefficientsl1} \\
\beta_{l2}(k)&=&\alpha_l(k)
	\frac{\hat{j}_l(ka)\hat{h}_l^{+\prime}(Qa)-
 \frac{k}{Q}\hat{j}_l^{\prime}(ka)\hat{h}_l^+(Qa)}
	{\hat{h}_l^-(Qa)\hat{h}_l^{+\prime}(Qa)-
 \hat{h}_l^{-\prime}(Qa)\hat{h}_l^+(Qa)} \\
{\cal F}_{l1}(k)&=&
	\frac{[\hat{h}_l^+(Qb)\hat{h}_l^{-\prime}(kb)-
 \frac{Q}{k}\hat{h}_l^{+\prime}(Qb)\hat{h}_l^-(kb)]
	[\hat{j}_l(ka)\hat{h}_l^{-\prime}(Qa)-
 \frac{k}{Q}\hat{j}_l^{\prime}(ka)\hat{h}_l^-(Qa)]}
	{[\hat{h}_l^+(kb)\hat{h}_l^{-\prime}(kb)-
 \hat{h}_l^{+\prime}(kb)\hat{h}_l^-(kb)]
	[\hat{h}_l^+(Qa)\hat{h}_l^{-\prime}(Qa)-
 \hat{h}_l^{+\prime}(Qa)\hat{h}_l^-(Qa)]}\alpha_l(k) \nonumber \\
	&&{} +
 \frac{[\hat{h}_l^-(Qb)\hat{h}_l^{-\prime}(kb)-
 \frac{Q}{k}\hat{h}_l^{-\prime}(Qb)\hat{h}_l^-(kb)]
	[\hat{j}_l(ka)\hat{h}_l^{+\prime}(Qa)-
 \frac{k}{Q}\hat{j}_l^{\prime}(ka)\hat{h}_l^+(Qa)]}
	{[\hat{h}_l^+(kb)\hat{h}_l^{-\prime}(kb)-
 \hat{h}_l^{+\prime}(kb)\hat{h}_l^-(kb)]
	[\hat{h}_l^-(Qa)\hat{h}_l^{+\prime}(Qa)-
 \hat{h}_l^{-\prime}(Qa)\hat{h}_l^+(Qa)]}\alpha_l(k) 
 \nonumber \\
 && \quad \\
{\cal F}_{l2}(k)&=&
 \frac{[\hat{h}_l^+(Qb)\hat{h}_l^{+\prime}(kb)-
 \frac{Q}{k}\hat{h}_l^{+\prime}(Qb)\hat{h}_l^+(kb)]
	[\hat{j}_l(ka)\hat{h}_l^{-\prime}(Qa)-
 \frac{k}{Q}\hat{j}_l^{\prime}(ka)\hat{h}_l^-(Qa)]}
	{[\hat{h}_l^-(kb)\hat{h}_l^{+\prime}(kb)-
 \hat{h}_l^{-\prime}(kb)\hat{h}_l^+(kb)]
	[\hat{h}_l^+(Qa)\hat{h}_l^{-\prime}(Qa)-
 \hat{h}_l^{+\prime}(Qa)\hat{h}_l^-(Qa)]}\alpha_l(k) \nonumber \\
	&&{} +
 \frac{[\hat{h}_l^-(Qb)\hat{h}_l^{+\prime}(kb)-
 \frac{Q}{k}\hat{h}_l^{-\prime}(Qb)\hat{h}_l^+(kb)]
	[\hat{j}_l(ka)\hat{h}_l^{+\prime}(Qa)-
 \frac{k}{Q}\hat{j}_l^{\prime}(ka)\hat{h}_l^+(Qa)]}
	{[\hat{h}_l^-(kb)\hat{h}_l^{+\prime}(kb)-
 \hat{h}_l^{-\prime}(kb)\hat{h}_l^+(kb)]
	[\hat{h}_l^-(Qa)\hat{h}_l^{+\prime}(Qa)-
 \hat{h}_l^{-\prime}(Qa)\hat{h}_l^+(Qa)]}\alpha_l(k) \, .
 \nonumber \\ 
\end{eqnarray}
\end{mathletters} 
The $l$th radial ket reads up to a normalization factor as
\begin{equation}
 \chi_l(r;E)=\left\{ \begin{array}{ll}
 \alpha_l(k)\hat{j}_l(kr) &0<r<a \\
 \alpha_{l2}(k)\hat{h}_l^+(Qr)
 +\beta_{l2}(k)\hat{h}_l^-(Qr) &a<r<b \\
 {\cal F}_{l1}(k)\hat{h}_l^+(kr)+
 {\cal F}_{l2}(k)\hat{h}_l^-(kr) 
 &b<r<\infty \, .
 \end{array} 
 \right. \label{RLDkkert}
\end{equation}

\subsection{S-Matrix Approach}

Now we calculate the $l$th partial $S$ matrix and its poles.
Because 
\begin{equation}
 \hat{h}_l^{\pm}\sim e^{\pm i (kr-l\pi /2)} \, , \qquad r \to
\infty \, ,
	\label{ae1}
\end{equation}
the function ${\cal F}_{l1}\hat{h}_l^+$ may be interpreted as an
outgoing wave with amplitude ${\cal F}_{l1}$ and ${\cal
F}_{l2}\hat{h}_l^-$ may be interpreted as an incoming wave with
amplitude ${\cal F}_{l2}$. Thus the expression for the $l$th
partial $S$ matrix in the energy representation is
\begin{equation}
	S_l(k)=-\frac{{\cal F}_{l1}(k)}{{\cal F}_{l2}(k)} \, .
 \label{lsmatrix}
\end{equation}
The resonances are associated with the poles of the analytic 
continuation of Eq.~(\ref{lsmatrix}). The functions ${\cal F}_{l1}$
and
${\cal F}_{l2}$ are analytic, and therefore the 
poles of the $S_l$ matrix (\ref{lsmatrix}) are the zeros of its 
denominator.\cite{MARSDEN} Then, the condition
${\cal F}_{l2}(k)=0$ provides the $l$th resonance energies,
\begin{eqnarray}
 &&[\hat{h}_l^+(Qb)\hat{h}_l^{+\prime}(kb)-
 \frac{Q}{k}\hat{h}_l^{+\prime}(Qb)\hat{h}_l^+(kb)]
	[\hat{j}_l(ka)\hat{h}_l^{-\prime}(Qa)-
 \frac{k}{Q}\hat{j}_l^{\prime}(ka)\hat{h}_l^-(Qa)] \nonumber \\
&&{}-[\hat{h}_l^-(Qb)\hat{h}_l^{+\prime}(kb)-
 \frac{Q}{k}\hat{h}_l^{-\prime}(Qb)\hat{h}_l^+(kb)]
	[\hat{j}_l(ka)\hat{h}_l^{+\prime}(Qa)-
 \frac{k}{Q}\hat{j}_l^{\prime}(ka)\hat{h}_l^+(Qa)] =0 
 \, . \nonumber \\
 \label{resonancesl}
\end{eqnarray}

\subsection{Gamow Vector Approach}
As we did for $l=0$, we shall prove that the $S$-matrix poles are the same as
the complex eigenvalues obtained from the purely outgoing boundary condition.

The Gamow vector in the position representation satisfies
\begin{equation}
	\left(\frac{-\hbar^2}{2m}\frac{1}{r}\frac{\partial^2}{\partial r^2}r
	+\frac{\hbar^2l(l+1)}{2mr^2}+V(r)\right)
 \langle r,\theta, \phi|z_R,l,m \rangle 
 = z_R\langle r,\theta, \phi|z_R,l,m \rangle \, .
	\label{GsphSe}
\end{equation}
If we write
$\langle r,\theta,\phi|z_R,l,m\rangle:=
\frac{1}{r}\chi_l(r;z_R)Y_{l,m}(\theta,\phi )$, then the radial part fulfills
\begin{equation}
	\left(\frac{-\hbar^2}{2m} \frac{d^2}{dr^2}+
	\frac{\hbar^2l(l+1)}{2mr^2}
	+V(r) \right) \chi_l(r;z_R)=z_R \chi_l(r;z_R) \, .
 \label{RlGV}
\end{equation}
The general solution of (\ref{RlGV}) is 
\begin{equation}
 \chi_l(r;z_R)=\left\{ \begin{array}{ll} 
 \alpha_{l1} \hat{h}_l^+(kr)+\beta_{l1} \hat{h}_l^-(kr) & 0<r<a \\
	\alpha_{l2} \hat{h}_l^+(Qr)+\beta_{l2} \hat{h}_l^-(Qr) & a<r<b \\
	{\cal F}_{l1} \hat{h}_l^+(kr)+{\cal F}_{l2} \hat{h}_l^-(kr) 
 & b<r<\infty \, ,
 \end{array} \right.
\end{equation}
where $k$ is the complex wave number (\ref{cmomentum}) and $Q$ is
given by Eq.~(\ref{Qcmomentum}). The boundary conditions that the
Gamow vectors satisfy are equivalent to the case $l=0$. Now
$\hat{h}_l^+$ plays the role of outgoing wave and
$\hat{h}_l^-$ that of the incoming wave,
\begin{mathletters}
 \label{Gbcl}
\begin{eqnarray}
	\chi_l(0;z_R)&=&0 \\
	\chi_l(a-;z_R)&=&\chi_l(a+;z_R) \\
	\chi_l'(a-;z_R)&=&\chi_l'(a+;z_R) \\
	\chi_l(b-;z_R)&=&\chi_l(b+;z_R) \\
 \chi_l'(b-;z_R)&=&\chi_l'(b+;z_R) \\ 
	\chi_l(r;z_R)&\sim& \hat{h}_l^+(kr) \, ,  \qquad r\to \infty \, .
\end{eqnarray}
\end{mathletters} 
If we define $\alpha_l=2i\alpha_{l1}$, Eq.~(\ref{Gbcl}) can be written 
in terms of the coefficients of $\chi_l(r;z_R)$ as
\begin{mathletters}
\label{eq:mat}
\begin{eqnarray}
	 \alpha_{l2} \hat{h}_l^+(Qa)+\beta_{l2} \hat{h}_l^-(Qa)&=&
 \alpha_l \hat{j}_l(ka) \\
	\alpha_{l2}\hat{h}_l^{+\prime}(Qa)+
 \beta_{l2}\hat{h}_l^{-\prime}(Qa)&=&
 \frac{k}{Q}\alpha_l \hat{j}_l^{\prime}(ka) \\
	{\cal F}_{l1} \hat{h}_l^+(kb)&=&
	\alpha_{l2}\hat{h}_l^+(Qb)+\beta_{l2} \hat{h}_l^-(Qb) \\
	{\cal F}_{l1}\hat{h}_l^{+\prime}(kb)&=&
	\frac{Q}{k}(\alpha_{l2}\hat{h}_l^{+\prime}(Qb)+
 \beta_{l2}\hat{h}_l^{-\prime}(Qb)) \, .
\end{eqnarray}
\end{mathletters}
In the matrix representation Eq.~(\ref{eq:mat}) looks like
\begin{equation}
	\left(\begin{array}{cccc} 
 -\hat{j}_l(ka)&0&\hat{h}_l^+(Qa)&\hat{h}_l^-(Qa)\\
 \frac{-k}{Q}\hat{j}_l^{\prime}(ka)&0&\hat{h}_l^{+\prime}(Qa)&
 \hat{h}_l^{-\prime}(Qa)\\
 0&-\hat{h}_l^+(kb)&\hat{h}_l^+(Qb)&\hat{h}_l^-(Qb)\\
 0&-\hat{h}_l^{+\prime}(kb)&\frac{Q}{k}\hat{h}_l^{+\prime}(Qb)&
 \frac{Q}{k}\hat{h}_l^{-\prime}(Qb)
 \end{array}\right)
 \left(\begin{array}{cccc} \alpha_l \\ {\cal F}_{l1}\\ \alpha_{l2} \\
 \beta_{l2}
 \end{array}\right)=
 \left(\begin{array}{cccc} 0\\ 0\\0 \\ 0
 \end{array}\right) \, .
\end{equation}
This system has non-trivial solutions iff the determinant of the 
coefficients is equal to zero, 
\begin{equation}
	\left|\begin{array}{cccc} 
 -\hat{j}_l(ka)&0&\hat{h}_l^+(Qa)&\hat{h}_l^-(Qa)\\
 \frac{-k}{Q}\hat{j}^{\prime}_l(ka)&0&\hat{h}_l^{+\prime}(Qa)&
 \hat{h}_l^{-\prime}(Qa)\\
 0&-\hat{h}_l^+(kb)&\hat{h}_l^+(Qb)&\hat{h}_l^-(Qb)\\
 0&-\hat{h}_l^{+\prime}(kb)&\frac{Q}{k}\hat{h}_l^{+\prime}(Qb)&
 \frac{Q}{k}\hat{h}_l^{-\prime}(Qb)
 \end{array} \right|=0 \, .
\end{equation}
Straightforward calculations then lead to
\begin{eqnarray}
\label{eq:mat2}
 &&[\hat{h}_l^+(Qb)\hat{h}_l^{+\prime}(kb)-
 \frac{Q}{k}\hat{h}_l^{+\prime}(Qb)\hat{h}_l^+(kb)]
	[\hat{j}_l(ka)\hat{h}_l^{-\prime}(Qa)-
 \frac{k}{Q}\hat{j}_l^{\prime}(ka)\hat{h}_l^-(Qa)] \nonumber \\
 &&{}-[\hat{h}_l^-(Qb)\hat{h}_l^{+\prime}(kb)-
 \frac{Q}{k}\hat{h}_l^{-\prime}(Qb)\hat{h}_l^+(kb)]
	[\hat{j}_l(ka)\hat{h}_l^{+\prime}(Qa)-
 \frac{k}{Q}\hat{j}^{\prime}_l(ka)\hat{h}_l^+(Qa)] =0 
 \, .\nonumber \\
 \quad
\end{eqnarray}
Equation~(\ref{eq:mat2}) is just Eq.~(\ref{resonancesl}), which
was found in the previous section using the $S$ matrix approach.

\subsection{Green Function Approach}
The $l$th radial Green function satisfies
\begin{equation}
 \left( -\frac{\hbar^2}{2m}\frac{\partial^2}{\partial r^2}
 +\frac{\hbar^2l(l+1)}{2mr^2}+V(r)-E\right)
 G_l(r,r';E)=-\delta (r-r') \, .
 \label{lgreene}
\end{equation}
It expression is given by
\begin{equation}
 G_l(r,r';E)=
 \frac{2m}{\hbar^2} \, 
 \frac{\chi_l(r_<;E) \psi_l(r_>;E)}{W(\chi_l ,\psi_l )} \, .
\end{equation}
The function $\chi_l$ is the solution of the time independent 
Schr\"odinger equation~(\ref{baba}) that vanishes at the origin. Therefore,
$\chi_l(r;E)$ is given by Eq.~(\ref{RLDkkert}). The function
$\psi_l$ also satisfies Eq.~(\ref{baba}), but with the
boundary conditions
\begin{mathletters}
      \label{bcpsisis}
\begin{eqnarray}
	\psi_l(a-)&=&\psi_l(a+) \\
	\psi_l'(a-)&=&\psi_l'(a+) \\
	\psi_l(b-)&=&\psi_l(b+) \\
	\psi_l'(b-)&=&\psi_l'(b+) \\
	\psi_l(r)&\sim & \hat{h}_l^+(kr) \, , \qquad r\to \infty \, .
\end{eqnarray}
\end{mathletters} 
Then
\begin{equation}
 \psi_l(r;k)=\left\{ \begin{array}{ll}
 a_{l1}(k)\hat{h}_l^+(kr)+
 b_{l1}(k)\hat{h}_l^-(kr) &0<r<a \\
 a_{l2}(k)\hat{h}_l^+(Qr)
 +b_{l2}(k)\hat{h}_l^+(Qr) &a<r<b \\
 \hat{h}_l^+(kr) &b<r<\infty \, ,
 \end{array} 
 \right.
\end{equation}
where the $a$ and $b$ coefficients express the continuity 
conditions in Eq.~(\ref{bcpsisis}). The $l$th Green function is given by
\begin{equation}
 G_l(r,r';k)=
 \frac{2m}{\hbar^2}\, 
 \frac{\chi_l(r_<;k) \, \psi_l(r_>;k)}{2ik{\cal F}_{2l}(k)} \, .
 \label{Glgresj}
\end{equation}
The poles of Eq.~(\ref{Glgresj}) are the zeros of its 
denominator.\cite{MARSDEN} This is the same condition as was 
found in the $S$-matrix approach and leads to the same result as the Gamow
vector approach.

\newpage

\newpage

\newpage

\begin{thebibliography}{99}

\bibitem{BARNETT}D.~E.~Groom {\it et al.}, ``Review of 
particle physics,'' Eur.~Phys.~J.~C{\bf 15}, 1--878
(2000). 

\bibitem{WW} V.~E.~Weisskopf and E.~P.~Wigner, ``Berechnung der
nat\"urlichen Linienbreite auf Grund der Diracsichen Licht
theory,'' Z.~Phys.~{\bf 63}, 54--73 (1930); ``Uber die
nat\"urlichen Linienbreite in der Strahlung des harmonischen
Oszillators,'' Z.~Phys.~{\bf 65}, 18--27 (1930). 

\bibitem{BOHM} A.~Bohm, {\it Quantum Mechanics: Foundations and 
Applications} (Springer-Verlag, New York, 1986). 

\bibitem{GAMOW} G.~Gamow, ``Zur Quantentheorie de Atomkernes,'' 
Z.~Phys.~{\bf 51}, 204--212 (1928). 

\bibitem{KPS} A.~F.~J.~Siegert, ``On the derivation of the
dispersion formula for nuclear reactions,'' Phys.~Rev.~{\bf 56},
750--752 (1939). 

\bibitem{PEI} R.~E.~Peierls, ``Complex eigenvalues in scattering
theory,'' Proc.~R.~Soc.~London, Ser.~A {\bf 253}, 16--36 (1959).

\bibitem{MONDRAGON} E.~Hern\'andez and A.~Mondrag\'on,
``Resonant states in momentum representation,'' Phys.~Rev.~C {\bf
29}, 722--738 (1984); A.~Mondrag\'on, E.~Hern\'andez, and
J.~M.~Vel\'azquez Arcos, ``Resonances and Gamow states in
non-local potentials,'' Ann.~Phys.~(Leipzig) {\bf 48}, 503--517
(1991).

\bibitem{FERREIRA} M.~Baldo, L.~S.~Ferreira, and L.~Streit, 
``Eigenvalue problem for Gamow vectors and a separable
approximation for the N-N interaction,'' Phys.~Rev.~C~{\bf 36},
1743--1746 (1987). L.~S.~Ferreira, E.~Maglione, R.~J.~Liotta,
``Nucleon resonances in deformed nuclei,'' Phys.~Rev.~Lett.~{\bf
78}, 1640--1643 (1997).

\bibitem{BOLLINI} C.~G.~Bollini, O.~Civitarese, A.~L.~De Paoli,
and M.~C.~Rocca, ``Gamow states as continuous linear functionals 
over analytical test functions,'' J.~Math.~Phys.~{\bf 37},
4235--4242 (1996). 

\bibitem{GASTON} G.~Garc\'\i a-Calder\'on and R.~Peierls, ``Resonant
states and their uses,'' Nucl.~Phys. A {\bf 265}, 443--460 (1976). 

\bibitem{BERGGREN} T.~Berggren, ``On the use of resonant states in 
eigenfunction expansions of scattering and reaction amplitudes,''
Nucl.~Phys.~A {\bf 109}, 265--287 (1968).

\bibitem{ROLLEFSON} A.~A.~Rollesfon and R.~M.~Prior, ``An advanced
undergraduate nuclear lifetime experiment,'' Am.~J.~Phys.~{\bf 46},
1007--1008 (1978).

\bibitem{SUPON} F.~W.~Supon and J.~J.~Kraushaar, ``Radioactive
half-life measurements in a freshman or sophomore laboratory,'' 
Am.~J.~Phys.~{\bf 51}, 761--763 (1983).

\bibitem{DESMARAIS} D.~Desmarais and J.~L.~Duggan, ``The study of
alpha-particle decay schemes of heavy nuclei,'' Am.~J.~Phys.~{\bf
58}, 1079--1085 (1990).

\bibitem{RUDDICK} K.~Ruddick, ``Determination of the half-life of
$^{212}$Po,'' Am.~J.~Phys.~{\bf 63}, 658--660 (1995).

\bibitem{DIS} R.~de la Madrid, {\it Quantum mechanics in rigged
Hilbert space language}, Ph.D.\ thesis, Universidad de Valladolid,
Valladolid, 2001.  Available at
\texttt{http://www.isi.it/$\sim$rafa/}.

\bibitem{ANTOINE98} J.-P.~Antoine, ``Quantum mechanics beyond
Hilbert space,'' in {\it Irreversibility and Causality}, 
A.~Bohm, H.-D.~Doebner, and P.~Kielanowski, eds. (Springer-Verlag,
Berlin, 1998), pp. 3--33.

\bibitem{BG} A.~Bohm and M.~Gadella, {\it Dirac kets, Gamow Vectors,
and Gelfand Triplets}, Springer Lectures Notes in Physics Vol.~348
(Springer, Berlin, 1989).

\bibitem{RHS1}J.-P.~Antoine, ``Dirac formalism and symmetry
problems in quantum mechanics. I. General Dirac formalism, II.
Dirac formalism and symmetry problems in quantum mechanics,''
J.~Math.~Phys.~{\bf 10}, 53--69, 2276--2290 (1969).

\bibitem{RHS2} A.~Bohm, ``The Rigged Hilbert Space in Quantum
Mechanics,'' 
\emph{Boulder Lectures in Theoretical Physics, 1966}, Vol. 9A
(Gordon and Breach, New York, 1967). 

\bibitem{RHS3} J.~E.~Roberts, ``The Dirac bra and ket formalism,''
J.~Math.~Phys.~{\bf 7}, 1097--1104 (1966); ibid., ``Rigged Hilbert
spaces in quantum mechanics,'' Commun.~Math.~Phys.~{\bf 3}, 98--119
(1966).

\bibitem{BGM} A.~Bohm, M.~Gadella, and G.~B.~Mainland, ``Gamow
vectors and decaying states,'' Am.~J.~Phys.~{\bf 57}, 1103--1108
(1989). 

\bibitem{HOLSTEIN95} B.~R.~Holstein, ``Understanding alpha decay,'' 
Am.~J.~Phys.~{\bf 64}, 1061--1071 (1996).

\bibitem{FUDA} M.~G.~Fuda, ``Time-dependent theory of alpha
decay,'' Am.~J.~Phys.~{\bf 52}, 838--842 (1984).

\bibitem{HOLSTEIN83} B.~R.~Holstein, ``Bound states, virtual
states, and nonexponential decay via path integrals,'' 
Am.~J.~Phys.~{\bf 51}, 897--901 (1983).

\bibitem{MASSMANN} H.~Massmann, ``Illustration of resonances and
the law of exponential decay in a simple quantum-mechanical
problem,'' Am.~J.~Phys.~{\bf 53}, 679--683 (1985).

\bibitem{LEVITAN} H.~Jakobovits, Y.~Rothschild, and J.~Levitan,
``The approximation to the exponential decay law,''
Am.~J.~Phys.~{\bf 63}, 439--443 (1995).

\bibitem{ONLEY} D.~S.~Onley and A.~Kumar, ``Time dependence in
quantum mechanics study of a simple decaying system,''
Am.~J.~Phys.~{\bf 60}, 432--439 (1992).

\bibitem{VON} J.~von Neumann, {\it Mathematische Grundlagen der 
Quantertheorie} (Springer, Berlin, 1931); English translation {\it
Mathematical Foundations of Quantum Mechanics} (Princeton
University Press, Princeton, 1955).

\bibitem{DIRAC} P.~A.~M.~Dirac, {\it The Principles of Quantum
Mechanics} (Clarendon Press, Oxford, 1958).

\bibitem{GELFAND} I.~M.~Gelfand and N.~Y.~Vilenkin, 
{\it Generalized Functions}, Vol.~IV (Academic Press, New York,
1964); K.~Maurin, {\it Generalized Eigenfunction Expansions and
Unitary Representations of Topological Groups} (Polish Scientific
Publishers, Warsaw, 1968). 

\bibitem{FP} R.~de la Madrid, A.~Bohm, and M.~Gadella, ``Rigged
Hilbert space treatment of continuous spectrum,''
Fortsch.~Phys.~{\bf 50}, 185--216 (2002); {\sf quant-ph/0109154}.

\bibitem{JPA} R.~de la Madrid, ``Rigged Hilbert space approach to
the Schr\"odinger equation,'' J.~Phys.~A: Math.~Gen.~{\bf 35}, 319--342 
(2002); {\sf quant-ph/0110165}. 

\bibitem{COHEN} C.~Cohen-Tannoudji, B.~Diu, and F.~Lalo\"e, 
{\it Quantum Mechanics} (Wiley, New York, 1977), Vol. II,
Chap.~VIII. 

\bibitem{DUNFORD} N.~Dunford and J.~Schwartz,
{\it Linear Operators} (Interscience Publishers, New York,
1963), Vol.~II.

\bibitem{NEWTON} R.~G.~Newton, {\it Scattering Theory of Waves and
Particles} (McGraw-Hill, New York, 1966); ibid. (Springer-Verlag,
New York, 1982), 2nd ed.

\bibitem{TAYLOR} J.~R.~Taylor, {\it Scattering Theory} (Wiley, New
York, 1972).

\bibitem{KUKULIN} V.~I.~Kukulin, V.~M.~Krasnopol'sky, and
J.~Horacek, {\it Theory of Resonances} (Kluwer Academic Publishers,
Dordrecht, 1989).

\bibitem{NUSSENZVEIG} H.~M.~Nussenzveig, {\it Causality and
Dispersion Relations} (Academic Press, New York, 1972).

\bibitem{MARSDEN} J.~E.~Marsden, {\it Basic Complex Analysis}
(Freeman, San Francisco, 1975).

\bibitem{CSF} R.~de la Madrid, ``Formal and precise derivation of
the Green functions for a simple potential,'' Chaos, Solitons and
Fractals~{\bf 12}, 2689--2695 (2001); {\sf quant-ph/0107096}.

\bibitem{GADELLA} M.~Gadella, ``Derivation of Gamow vectors for
resonances in cut-off potentials,'' Lett.~Math.~Phys.~{\bf 41},
279--290 (1997). 

\bibitem{BORN} M.~Born and E.~Wolf, 
{\it Principles of Optics: Electromagnetic Theory of Propagation,
Interference and Diffraction of Light} (Pergamon Press, 1980);
J.~Jackson, {\it Classical Electrodynamics} (John
Wiley \& Sons, New York, 1962), 2nd ed.
 
\bibitem{KHALFIN} L.~A.~Khalfin, ``Contribution to the decay theory
of a quasi-stationary state,'' Sov.~Phys.~JETP {\bf 6}, 1053--1063
(1958).

\bibitem{WATSON} G.~N.~Watson, {\it Theory of Bessel Functions}
(Cambridge University Press, Cambridge, 1958).

\end{thebibliography}
\end{document}